\documentclass[preprint,secnumarabic,amsmath,amssymb]{revtex4}
\usepackage{amsmath}

\newtheorem{remark}{Remark}[section]
\usepackage{graphicx}
\usepackage{epsf}
\usepackage{bm}
\usepackage{amsmath}
\usepackage{amsfonts}
\usepackage{amssymb}
\usepackage{epstopdf}
\usepackage{natbib}
\usepackage{color}
\usepackage[font=small,skip=2pt]{caption}
\usepackage{float}
\usepackage{titlesec}
\usepackage{titletoc}
\usepackage[title,titletoc]{appendix}
\renewcommand{\appendixname}{Appendix}

\begin{document}
\tolerance=5000
\noindent

\title{The spectrum of Gravitational Waves, their overproduction in   quintessential inflation and its influence in the reheating temperature}

\author{Jaume Haro$^{1}$\footnote{E-mail: jaime.haro@upc.edu}}

\affiliation{$^{1}$Departament de Matem\`atiques, Universitat Polit\`ecnica de Catalunya, Diagonal 647, 08028 Barcelona, Spain}

\author{Llibert Arest\'e Sal\'o$^{2}$\footnote{E-mail: llibert.areste-salo@tum.de}}

\affiliation{$^{2}$TUM Physik-Department, Technische Universit{\"a}t M{\"u}nchen, James-Franck-Str.1, 85748 Garching, Germany}




\begin{abstract}
One of the most important issues in an inflationary theory as standard or quintessential inflation is the  mechanism to reheat the universe after the end of the inflationary period in order to match with the Hot Big Bang universe. In quintessential inflation two mechanisms are frequently used, namely the reheating via gravitational particle production which is, as we will see, very efficient when the phase transition from the end of inflation to a kinetic regime (all the energy of the inflaton field is kinetic) is very abrupt, and the so-called {\it instant preheating} which is used for a very smooth phase transition because in that case the gravitational particle production is very inefficient.

\

In the present work, a detailed study of these mechanisms is done, obtaining bounds for the reheating temperature and the range of the parameters involved in each reheating mechanism in order that the Gravitational Waves (GWs) produced at the beginning of kination do not disturb the Big Bang Nucleosynthesis (BBN) success.

\vspace{0.5cm}


\end{abstract}

\vspace{0.5cm}



\maketitle

\section{ Introduction}

Soon after the discovery of the current cosmic acceleration at the end of the last century \cite{riess, perlmutter}, a class of pioneering cosmological models attempting to unify the early- and late- accelerating expansions were introduced. By construction these models, named as  {\it quintessential inflation} models \cite{Spokoiny, pr, pv}, 
unlike the standard quintessence ones  (see \cite{Tsujikawa} for a review of these models), 
 only contain one classical scalar field, also named inflaton as in standard inflation \cite{guth, linde, starobinsky, albrecht}, and it is shown that they succeed 
 in reproducing these two accelerated epochs of the universe (also see \cite{deHaro:2016hpl,hap,deHaro:2016hsh,deHaro:2016ftq,deHaro:2017nui,Geng:2017mic,AresteSalo:2017lkv, Haro:2015ljc, hossain1,hossain2,hossain3,hossain4,guendelman1} for other interesting {\it quintessential inflation} models).

\

However, an important difference occurs with respect to the standard inflationary paradigm, where 
the potential of the inflaton field has a local minimum and, thus, the inflaton field  releases its energy while it oscillates, which allows particle production \cite{kls, kls1, gkls, stb, Basset}. 
In contrast, for the  ``non oscillating"  models, i.e. in {\it quintessential inflation},  where the inflation field survives to be able to reproduce the current cosmic acceleration, a fast phase transition from the end of inflation to the beginning of kination (a regime where all the energy density of the inflation field is kinetic) where the adiabatic regime is broken 
is needed in order to reheat the universe. This creates an enough amount of particles which, after decays and/or  interactions with other fields, form a thermal relativistic plasma whose energy density will eventually become dominant.  The mechanism of particle creation can be obtained in different ways, 
but the most used and the ones we will study in this work  are 
 the  gravitational particle production \cite{Parker,gm,glm,gmm, ford, Birrell1, Zeldovich, dimopoulos0, hashiba} and the  {\it instant preheating}
\cite{fkl0, fkl, dimopoulos, vardayan}  (see also \cite{haro18} for a detailed description of both mechanisms).

\

Dealing with the mechanisms to reheat the universe, the question related to the bounds of the reheating temperature arises.
Some works have already considered the constraints for reheating in quintessential inflation models, both on {\it instant preheating} \cite{sami} and on gravitational particle production \cite{figueroa}.
A lower bound is obtained recalling  that the
radiation dominated era is prior to the Big Bang  Nucleosynthesis (BBN) epoch which occurs in the  $1$ MeV  regime \cite{gkr}. 
As a consequence,  the reheating temperature has to be greater
than $1$ MeV {(see also \cite{hasegawa}   where the authors obtain lower limits on the reheating temperature in the MeV  regime assuming both radiative and hadronic decays of relic particles only gravitationally interacting and taking into account effects of neutrino self-interactions and oscillations   in the neutrino thermalization calculations.)}  The upper bounds may depend on the theory we are dealing with; for instance, many supergravity and superstring theories contain particles such as the gravitino or a modulus field with only gravitational interactions and, thus,
the late time decay of these relic products may disturb  
the success of the standard BBN \cite{lindley}, but
this problem can be successfully removed if the  reheating temperature is
of the order of $10^9$ GeV (see for instance \cite{eln}).  This is the reason why  we will restrict  the reheating temperature to remain, more or less,  between $1$ MeV and $10^9$ GeV.

\

On the other hand, one has to take into account that a viable reheating mechanism has to deal with the affectation of the Gravitational Waves (GWs) in the BBN success by satisfying the observational bounds coming from the overproduction of the GWs \cite{pv} or related to the logarithmic spectrum of its energy density \cite{maggiore}.
As we will see throughout this work, the overproduction of GWs constrains very much the value of the parameters involved in the different reheating mechanisms and also impose hard bounds in the reheating temperature.

\

{In addition, another issue related to quintessential inflation is the possibility to explain the present abundance of dark matter.  Effectively,  assuming that
dark matter is made of non-decaying superheavy particles only coupled to gravity which are gravitationally created during the abrupt phase transition, one can show that a certain range of mass values of the dark matter leads to a viable model overpassing all the bounds coming from the overproduction of GWs \cite{haro19,ha19}.

\

 The manuscript is organized as follows:
In Section II we deduce the initial condition to apply the WKB approximation and ensure that the vacuum fluctuation of a massive field coupled to gravity  does not affect the classical evolution of the inflaton field. Section III is devoted to the presentation of our quintessential inflation  model, inspired in the well-known Peebles-Vilenkin one \cite{pv}, i.e., depending  on two parameters and  containing an abrupt phase transition from the end of inflation to the beginning of kination,  and the subsequent  study of its dynamical evolution. Next, in Section IV we study both  reheating mechanisms in quintessential inflation, namely via gravitational particle production and via {\it instant preheating}, obtaining 
bounds for the reheating temperature. In Section V we deal with the constraints to preserve the BBN success coming from the logarithmic spectrum of GWs and also from its overproduction during the phase transition from the end of inflation to the beginning of kination, obtaining the range of values of the parameters involved in each reheating mechanism and   also more restrictive bounds for the reheating temperature. In Section VI we consider the present abundance of dark matter, assuming that it is composed by superheavy particles conformally coupled to gravity, which are also produced during the abrupt phase transition from the end of inflation to the beginning of kination, obtaining bounds for its mass. In Section VII we consider another quintessential inflation model with a more abrupt phase transition and we show the importance of this fact and the differences with the previous model. Finally, in the conclusions we discuss the obtained results.  

\

}

\section{Initial conditions for inflation and  the application of the   WKB approximation}
\label{sec-II}

{ We want to know when
one can apply the WKB solution in the early universe (see for instance \cite{Winitzki, Haro} in order to approximately find the modes of a field coupled to gravity.
This is very important because it allows us to compute analytically important quantities such as the vacuum polarization and the energy density of the produced particles after an abrupt phase transition. In order to do all the analytic calculations we will consider 
a potential like the one used by Peebles and Vilenkin in \cite{pv}
with a discontinuity in some derivative and, thus, 
we can obtain an analytic expression of the reheating temperature depending on the parameters involved in the reheating mechanism (the mass of the produced particles, the decay rate, the coupling constant between the quantum field which produces the particles, the inflaton field,...). 
}

\

 So, first at all it is well-known that at temperatures of the  order of the Planck's mass quantum effects become very important  and the classical picture of the universe is not possible. However, at temperatures below $M_{pl}$, for example 
at GUT scales (i.e., when the temperature is of the order of $T\sim 4\times 10^{-3} M_{pl}\sim  10^{16}$ GeV), the beginning of the  Hot Big Bang (HBB) scenario is possible. For the  flat FLRW universe the energy density of the universe, namely $\rho$,  and the Hubble parameter $H$ are related through $\rho=3H^2M_{pl}^2$, and, { for a universe filled with radiation},  the temperature of the universe is related to the energy density via $\rho = (\pi^2/30)g_{*} T^4$, where  the degrees of freedom for the Standard Model are
$g_*=106.75$ (see for instance \cite{rg}). Thus, one can conclude that a classical picture of the universe would be possible when $H\cong 5\times 10^{-5} M_{pl}\cong 10^{14}$ GeV. Now we consider that 
inflation starts at this scale, i.e., we take the value of the Hubble parameter at the beginning of inflation (denoted by $H_{beg}$) as $H_{beg}=5\times 10^{-5} M_{pl}$, and we assume that a quantum $\chi$-field coupled to gravity and/or to the inflaton field, which will be the responsible to reheat the universe, is in the vacuum at the beginning of inflation. If we choose
the mass of the  $\chi$-field at least one  order greater than this value of  the Hubble parameter ($m_{\chi}\geq H_{beg}\cong 5\times 10^{-4} M_{pl}\cong 10^{15}$ GeV, which is a mass of the same order as those of the vector mesons responsible for transforming quarks into leptons in simple theories with  SU(5)  symmetry \cite{lindebook}),  one can apply the WKB approximation to calculate
 the re-normalized  energy density of the vacuum. After subtracting the adiabatic modes up to order four, we obtain an energy density of the order   
 $H^6/m_{\chi}^2$ \cite{kaya}, which is subdominant compared to the energy density of the background $3H^2M_{pl}^2$ and, thus, does not affect  the  classical evolution of the inflation up to an abrupt phase transition where the adiabatic regime is broken, the $\chi$-field stops being in the  vacuum  and   particles are copiously produced with an energy density which decays slower than the one of the inflation, thus becoming eventually dominant.

  \

The dynamical evolution of the vacuum modes could be understood as follows: the $k$-vacuum mode  during the adiabatic regime can be approximated by ${\chi}_{k, WKB}^{(n)}$, where $n$ is the order of the WKB approximation, 
but, when
the adiabatic regime breaks down during a period of time, the WKB approximation cannot be used and only at the end of this period one can again use
it. But now the vacuum mode is a combination of positive and negative frequency modes which can be approximated by a linear combination of ${\chi}_{k,WKB}^{(n)}$ and its conjugate of  the form $\alpha_{k,n}{\chi}_{k, WKB}^{(n)} + \beta_{k,n}({\chi}_{k,WKB}^{(n)} )^*  $, where $\alpha$ and $\beta$ are the so-called {\it Bogoliubov coefficients}, and it is the manifestation  of the gravitational particle production. Basically this  is  the  viewpoint of  particle creation in curved space-times  \cite{Parker}, where the $\beta$-Bogoliubov coefficient, which is calculated matching  the modes before and after the discontinuity for models with a discontinuity in some derivative of the potential as the one introduced by Peebles-Vilenkin in \cite{pv}. This is the key ingredient to calculate the energy density of the produced particles.

In fact, the energy density of the produced particles after the end of the phase transition evolves as \cite{Birrell}
\begin{eqnarray}
 \rho_{\chi}(\tau)=\frac{1}{2\pi^2 a^4(\tau)}\int_0^{\infty} \omega_k(\tau)k^2|\beta_k|^2dk,
 \end{eqnarray}
 where $\omega_k(\tau)$ is the time dependent  frequency of the $k$-mode, and when $\beta_k$ is known we have an analytic expression of this energy density that allows us to calculate the reheating temperature and deduce its bounds.

  \

\section{The Peebles-Vilenkin model}
\label{sec-III}

In order to deal with an analytically solvable problem, i.e., having an analytic expression of the $\beta$-Bogoliubov coefficient, we consider a sudden phase transition where the third derivative of the Hubble parameter is discontinuous, which happens for the following {\it improved version} of the well-known Peebles-Vilenkin  quintessential inflationary potential \cite{pv},
\begin{eqnarray}\label{PV}
V(\varphi)=\left\{\begin{array}{ccc}
\lambda M_{pl}^4\left(1-e^{\sqrt{\frac{2}{3}}\frac{\varphi}{M_{pl}}}\right)^2 + \lambda M^4 & \mbox{for} & \varphi\leq 0\\
\lambda\frac{M^8}{\varphi^{4}+M^4} &\mbox{for} & \varphi\geq 0,\end{array}
\right.
\end{eqnarray}
where $\lambda$ is a dimensionless parameter and $M$ is a very small mass compared with the Planck one.

\

{ Here,}
 it is important to point out that
the inflationary { part} of the original Peebles-Vilenkin potential is a quartic potential and, thus, the theoretical values of the spectral index and the ratio of tensor to scalar perturbations do not enter in the marginalized  joint confidence contour in the plane  $(n_s,r)$ at $2\sigma$ CL \cite{Planck} without the presence of the running \cite{hap}. 
This is the reason why one has to change the quartic part  by a Starobinsky-type potential, whose spectral values do actually enter in this contour. 

\

{ The value of the parameter $\lambda$ is calculated as follows:}  we use the theoretical and observational values of the power spectrum  of the curvature fluctuation in a co-moving coordinate system when the pivot scale leaves the Hubble radius  \cite{btw},
${\mathcal P}_{\zeta}\cong \frac{H_*^2}{8\pi^2 M_{pl}^2\epsilon_*}\sim 2\times 10^{-9}$, where $\epsilon=-\frac{\dot{H}}{H^2}\cong\frac{M_{pl}^2}{2}\left(\frac{V_{\varphi}}{V}\right)^2$ is the main slow-roll parameter and the { star} ``$\ast$''  means that the quantity is evaluated when the pivot scale leaves the Hubble radius, obtaining
\begin{eqnarray}
\lambda\sim 9\pi^2(1-n_s)^2\times 10^{-9},
\end{eqnarray}
where we have used that for our model one has $\epsilon_*\cong \frac{3}{16}(1-n_s)^2$, 
  where $n_s$ denotes the spectral index
 and during inflation
$H_*^2\cong \frac{\lambda}{3}M_{pl}^2$. 

\

From the recent observations by Planck \cite{Planck} the value of the spectral index is constrained to be $n_s=0.968\pm 0.006$. Thus, taking its central value  one gets $\lambda\cong 9\times 10^{-11} $, which means that  $H_*\cong 5.4{ 8}\times 10^{-6} M_{pl}$. The tensor-to-scalar perturbation ration $r$ for this model yields $r=16\epsilon_*\approx 3(1-n_s)^2$, which leads for this range of values of $n_s$ to a small enough quantity ($r\leq 0.00581$ at $2\sigma$ C.L.) which agrees with the observational constraints.

\

{ On the other hand,  note that for our toy model the  second derivative of the potential is discontinuous  at $\varphi=0$, nearly at the beginning of the kination phase (In order to simplify, we will assume that kination starts when $\varphi=0$ because, as is shown in Figure \ref{retrat}, the maximum value of the kinetic energy is very close to $\varphi=0$). In addition, using Raychaudhuri equation, one can see that the third derivative of the Hubble rate is discontinuous at the beginning of kination, hence 
allowing particle production because the adiabatic evolution is broken. For example, if one considers a massive $\chi$-field coupled to gravity,
the fourth derivative of the frequency  $\omega_k(\tau)=\sqrt{k^2+a^2(\tau)m_{\chi}^2}$ is discontinuous for any $k$-mode. 

\

In fact, this kind of potentials with discontinuities was
studied by Starobinsky and others in  \cite{starobinsky0, starobinsky1}, who showed that the discontinuity of the effective potential could be obtained introducing a second scalar field coupled to the inflaton that experiences a cosmological second order phase transition (see for instance the introduction of Linde's book \cite{lindebook} for some simple examples of first and second order phase transitions), as is explained in Section 4 of \cite{starobinsky1} considering  the standard toy model used many times in the hybrid inflationary scenario \cite{lindehybrid}.

\

 What is important is that we have to understand the breakdown of the adiabatic behavior, at least for a more smooth potential, as follows:
 {\begin{eqnarray}\label{4} \frac{1}{\omega^5(\tau)}{\frac{d^4\omega_k(\tau)}{d\tau^4}}\geq 1 \end{eqnarray}}
in a region close 
to the beginning of kination with a characteristic time less than $\left(H(t)\right)^{-1}$ and, thus, in this region the adiabatic regime is broken, allowing the production of particles. Unfortunately, in this situation the analytic calculation of the energy density of the produced particles is not possible. This is the  reason why we  consider our toy model (\ref{PV}), where one can  get an analytic expression of this energy density (Note that the second derivative of   (\ref{PV})  is discontinuous, meaning that the third derivative of the Hubble rate is discontinuous at the beginning of kination, that is, the fourth derivative of $\omega_{k}(\tau)$ is discontinuous at that moment and, thus, the non-adiabatic  condition    (\ref{4})   is met).

\

Finally, numerical calculations {(namely event-driven integration with an ode RK78 integrator) \cite{haro19} } show that at the beginning of kination one has $H_{kin}\cong 1.44\times 10^{-6} M_{pl}$ and, thus,
the energy density of the background at the beginning of kination is given by $\rho_{\varphi, kin}\cong 6.26\times 10^{-12} M_{pl}^4$.

\

 \begin{figure}[H]
\includegraphics[width=0.5\textwidth]{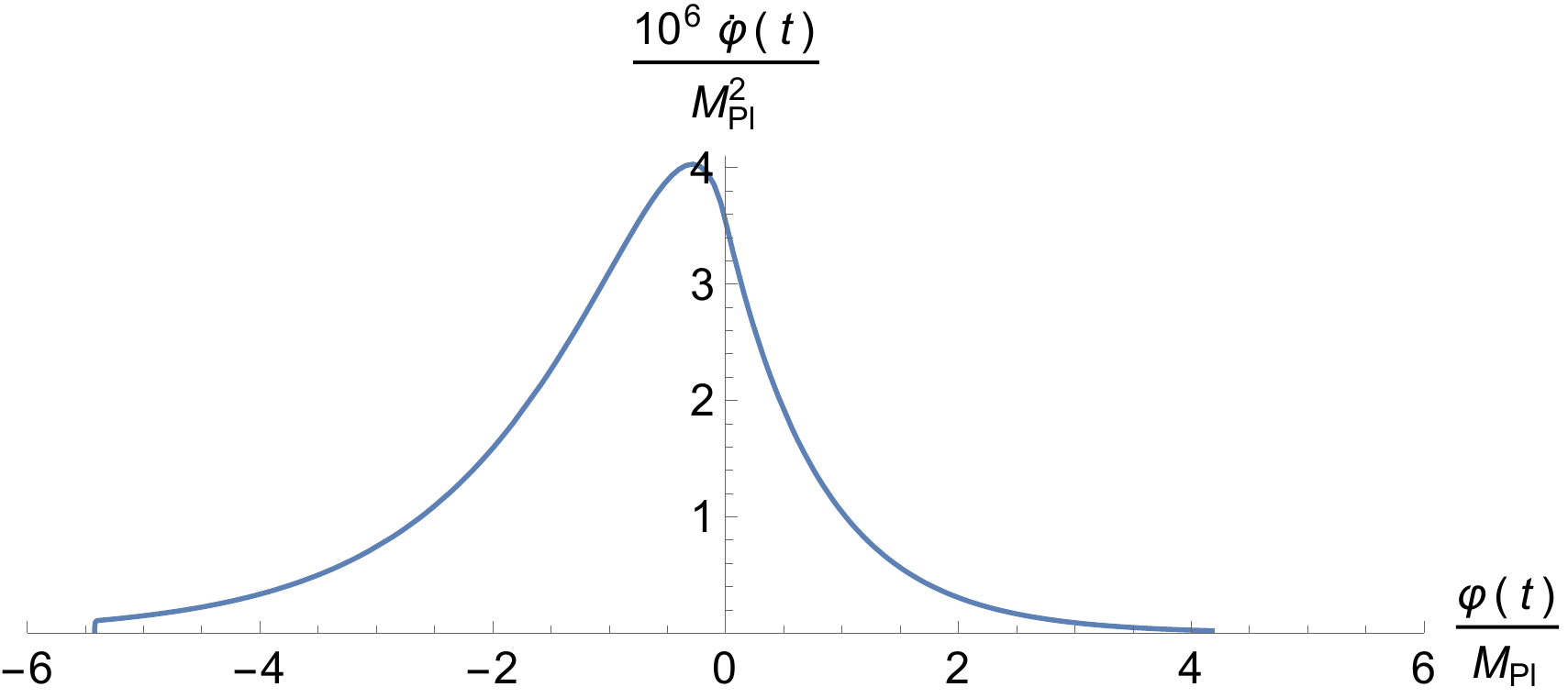}
\caption{Evolution of the velocity of the scalar field, as a function of scalar field,  obtained integrating the equation conservation $\ddot{\varphi}+3H\dot{\varphi}+V_{\varphi}=0$, with initial conditions when the pivot scale leaves the Hubble horizon, i.e., for
$\varphi_*=-5.42 M_{pl}$ and $\dot{\varphi}_*=0$. } 
\label{retrat}
\end{figure} 

\

\subsection{ The dynamics of the model}

To deal with the evolution of the system 
we need to  consider the back-reaction of particle production. Effectively,  after particle production one has the so-called
semi-classical Friedmann equation $H^2=\frac{\rho_{\varphi}+\rho_{\chi}}{3M_{pl}^2}$, where $\rho_{\varphi}$ stands for the energy density of produced particles and $\rho_{\chi}$ is the energy density of produced particles. Then, when the inflaton dominates, i.e., during inflation, kination and at late times,  one has 
$H^2=\frac{\rho_{\varphi}}{3M_{pl}^2}$, but from the end of kination up to beginning of quintessence one has $H^2=\frac{\rho_{\chi}}{3M_{pl}^2}$, that is, the background is driven by the energy density of created particles, meaning that during radiation one has $H=\frac{1}{3t}$ and in the matter-domination era $H=\frac{2}{3t}$, which influences the evolution of the infation field, which is given by the conservation equation $\ddot{\varphi}+3H\dot{\varphi}+V_{\varphi}=0$. Taking into account this fact, we start with the analytic analysis and after it we do the numerics.

\

\subsubsection{Analytic results} \label{dyn-analytical}

We start with the initial conditions at the beginning of kination for our improved version of the Peebles-Vilenkin model:
\begin{eqnarray}
\varphi_{kin}=0, \quad \dot{\varphi}_{kin}=3.54\times 10^{-6} M_{pl}^2.
\end{eqnarray}
During kination, the scale factor and the Hubble rate evolve as $a\propto t^{1/3}\Longrightarrow H=\frac{1}{3t}$ and, from the Friedmann equation, the evolution in this phase will be
\begin{eqnarray}
\frac{\dot{\varphi}^2}{2}=\frac{M_{pl}^2}{3t^2}
\Longrightarrow 
\varphi(t)=\sqrt{\frac{2}{3}}M_{pl}\ln\left( t/t_{kin} \right)=
\sqrt{\frac{2}{3}}M_{pl}\ln \left( \frac{H_{kin}}{H(t)} \right).\end{eqnarray}
Here two different situations can occur: The superheavy $\chi$-particles created during the phase transition from the end of inflation to the beginning of kination could decay

\begin{enumerate}
    \item After the end of kination.
    \item Before the end of kination.
\end{enumerate}

\

In the first case, 
at the end of kination one has 
 \begin{eqnarray}
\varphi_{end}=-\sqrt{\frac{2}{3}}M_{pl}\ln\left( \sqrt{2}\Theta \right), \quad
\dot{\varphi}_{end}=2\sqrt{3}M_{pl}H_{kin}\Theta,
\end{eqnarray}
where we have used the relation $H_{end}=\sqrt{2}H_{kin}\Theta$ (see Subsection \ref{grav-heavy} for the deduction), being 
$\Theta\equiv  \frac{\rho_{\chi, kin}}{\rho_{\varphi, kin}}$ (the ratio of the energy density of the $\chi$-field to the one of the inflaton at the beginning of the kination phase)
the so-called {\it heating efficiency} 
{\cite{rubio}}.  

\

During the period between $t_{end}$ and $t_R$ ($t_R$ denotes the reheating time, i.e., when the universe starts to be radiation-dominated), in the case that the $\chi$-particles were superheavy, the universe is matter-dominated and, thus, the Hubble parameter becomes $H=\frac{2}{3t}$. During this epoch, the gradient of the potential could also be disregarded,  hence the equation of the scalar field becomes $\ddot{\varphi}+\frac{2}{t}\dot{\varphi}=0$ and, thus, 
\begin{eqnarray}
\varphi(t)=\varphi_{end}+\sqrt{\frac{2}{3}}M_{pl}\left(1-\frac{t_{end}}{t}\right),
\end{eqnarray}
where we have used that $\dot{\varphi}(t)=-\dot{\varphi}_{end}\left( \frac{t}{t_{end}} \right)^2$ with $\dot{\varphi}_{end}=\sqrt{\frac{2}{3}}\frac{M_{pl}}{t_{end}}$. Then,
one gets 
\begin{eqnarray}
\varphi_{R}=
\varphi_{end}+\sqrt{\frac{2}{3}}M_{pl}\left(1-\frac{H_R}{2H_{end}}\right)=\varphi_{end}+
\sqrt{\frac{2}{3}}M_{pl}\left(1-\frac{\pi}{6}\sqrt{\frac{g_*}{10}}\frac{T_R^2}{H_{kin}M_{pl}\Theta}\right),
\end{eqnarray}
having employed that
$H_R^2=\frac{2\rho_{\varphi,R}}{3M_{pl}}$ with $\rho_{\varphi,R}=\frac{\pi^2}{30}g_*T_R^4 $
and we also have that
\begin{eqnarray}
\dot{\varphi}_{R}
=\frac{\sqrt{3}}{4}\frac{M_{pl}H_{R}^2}{H_{kin}\Theta}.
\end{eqnarray}

During the radiation period one can continue disregarding the potential and the dynamical equation becomes  
$\ddot{\varphi}+\frac{3}{2t}\dot{\varphi}=0$, whose solution is given by
\begin{eqnarray}
\varphi(t)=\varphi_{R}+2\dot{\varphi}_{R}t_{R}
\left(1-\sqrt{\frac{t_{R}}{t}}\right)
\end{eqnarray}
and, thus, 
since $\dot{\varphi}_{R}t_{R}= \frac{\pi}{6}\sqrt{\frac{ g_*}{30}}\frac{T^2_{R}}{H_{kin}\Theta}$ (being $T_R$ the reheating temperature)
 at the matter-radiation equality, one has
\begin{eqnarray}\label{eq0}
 \varphi_{eq} =\varphi_{R}+     \frac{\pi}{3}\sqrt{\frac{ g_*}{30}}\frac{T^2_{R}}{H_{kin}\Theta}\left(1-\sqrt{\frac{4H_{eq}}{3H_{R}}}\right) 
 =\varphi_{R}+ \frac{\pi}{3}\sqrt{\frac{ g_*}{30}}\frac{T^2_{R}}{H_{kin}\Theta} 
 \left[1-\frac{2T_{eq}}{\sqrt{3}T_{R}}\left( \frac{g_{eq}}{g_{*}} \right)^{\frac{1}{4}}\right] \nonumber \\
\cong \varphi_{R}+ \frac{\pi}{3}\sqrt{\frac{ g_*}{30}}\frac{T^2_{R}}{H_{kin}\Theta}\cong \varphi_{R}+\frac{2T^2_{R}}{H_{kin}\Theta},
  \end{eqnarray}
 where $g_{eq}\cong 3.36$ are the degrees of freedom at the matter-radiation equality and $T_{eq}$ is the temperature of the radiation at the matter-radiation equality, which is related to the energy
 density via the relation $\rho_{eq}=\frac{\pi^2}{15}g_{eq}T^4_{eq}\cong 8.8\times 10^{-1} \mbox{eV}^4$ and, thus, given by $T_{eq}\cong 7.9\times 10^{-10}$ GeV $\ll T_R$.
 In the same way,  
 \begin{align}\label{doteq}
\dot{\varphi}_{eq}=\dot{\varphi}_{R}\frac{t_{R}}{t_{eq}}
\sqrt{\frac{t_{R}}{t_{eq}}}=\left(\frac{16g_{eq}}{9g_*}
\right)^{3/4}\left(\frac{T_{eq}}{T_{R}}\right)^3 \dot{\varphi}_{R}
\cong 1.7\frac{T_{eq}^3T_{R}}{M_{pl} H_{kin}\Theta}.
\end{align} 

\begin{remark}
To obtain the value of $\rho_{eq}$, we have chosen as the value of the cosmic red-shift at the matter-radiation equality $z_{eq}\equiv -1+\frac{a_0}{a_{eq}}=3365$, the value of the ratio of the energy density of the matter to the critical energy density  at the present time equal to $\Omega_{matt,0}=0.308$ and the value of the Hubble rate at the present time equal to
$H_0=1.42\times 10^{-33}$ eV. Then, since $\rho_{matt,0}=3H_0^2M_{pl}^3\Omega_{matt,0}$, one finally gets
\begin{eqnarray}
\rho_{eq}=2\rho_{matt,0}(1+z_{eq})^3=8.8\times 10^{-1} \mbox{ eV}^4.
\end{eqnarray}
\end{remark}

\

In the second case, i.e., when the decay of the  $\chi$-particles is before the end of kination, which always happens when reheating is via {\it instant preheating}, the beginning  of the radiation era coincides with the end of kination. Thus,

\begin{eqnarray}
\varphi_R=\sqrt{\frac{2}{3}}M_{pl}\ln \left( \frac{H_{kin}}{H_R} \right)
\end{eqnarray}
and, taking into account that
$H_R=\sqrt{\frac{2}{3}}\frac{\sqrt{\rho_R}}{M_{pl}}=
\frac{\pi}{3}\sqrt{\frac{g_*}{5}}\frac{T_R^2}{M_{pl}},$
we get
\begin{eqnarray}
\varphi_{R}=\sqrt{\frac{2}{3}}M_{pl}\ln\left(
\frac{3}{\pi}\sqrt{\frac{5}{g_*}\frac{H_{kin}M_{pl}}
{T_R^2}} \right), 
\quad
\dot{\varphi}_{R}={\sqrt{6}}M_{pl}H_R=
\pi \sqrt{\frac{2g_*}{15}}{T_R^2}.
\end{eqnarray}

During the radiation era, disregarding once again the potential,   we will have
\begin{eqnarray}
\varphi(t)=\varphi_{R}+2\dot{\varphi}_{R}t_{R}
\left(1-\sqrt{\frac{t_{R}}{t}}\right),
\end{eqnarray}
but now
$\dot{\varphi}_Rt_R=\sqrt{\frac{2}{3}}M_{pl}$, meaning that
\begin{eqnarray}
\varphi_{eq}=\varphi_R+2\sqrt{\frac{2}{3}}M_{pl}\left(
1-\sqrt{\frac{2H_{eq}}{3H_R}}\right)=
\varphi_R+2\sqrt{\frac{2}{3}}M_{pl}\left(
1-\sqrt{\frac{2}{3}}\left(\frac{g_{eq}}{g_*} \right)^{1/4}\frac{T_{eq}}{T_R}\right)\cong\nonumber\\
\cong \varphi_R+2\sqrt{\frac{2}{3}}M_{pl}
\end{eqnarray}
given that $T_{eq}\ll T_R$, and
\begin{eqnarray}
\dot{\varphi}_{eq}=
\dot{\varphi}_R\left(\frac{t_R}{t_{eq}}\right)^{3/2}=
\frac{4\pi}{9}\sqrt{\frac{g_{eq}}{5}}
\left(\frac{g_{eq}}{g_*}\right)^{1/4}\frac{T_{eq}^3}
{T_R}.
\end{eqnarray}

\

After the matter-radiation equality the dynamical equations cannot be solved analytically and, thus, one needs to use numerical methods  to compute them. In order to do that, we need to use a ``time'' variable  that we choose to be minus the number of $e$-folds up to the present epoch, namely $N\equiv -\ln(1+z)=\ln\left( \frac{a}{a_0}\right)$. Now, using  the variable $N$,  one can recast the  energy density of radiation (the energy density of the decay products of the $\chi$-field which we continue denoting by $\rho_{\chi}$) and  matter respectively as
\begin{eqnarray}
 \rho_{\chi}(N)= \frac{\rho_{eq}}{2}e^{4(N_{eq}-N)} ,\qquad
 \rho_{matt}(N)=\frac{\rho_{eq}}{2}e^{3(N_{eq}-N)},
\end{eqnarray}
where  
$N_{eq}=-\ln(1+z_{eq})\cong -8.121$ is the value of $N$ at the matter-radiation equality.

\

In order to obtain the dynamical system for our model, we 
introduce the following dimensionless variables,
 \begin{eqnarray}
 x=\frac{\varphi}{M_{pl}}, \qquad y=\frac{\dot{\varphi}}{H_0 M_{pl}},
 \end{eqnarray}
 where once again $H_0\cong 1.42 \times 10^{-33}$ eV denotes the current value of the Hubble parameter. Now, using the variable
 $N = - \ln (1+z)$ defined above and also the conservation equation $\ddot{\varphi}+3H\dot{\varphi}+V_{\varphi}=0$, we will have the following  non-autonomous dynamical system \cite{hap19}:
 \begin{eqnarray}\label{system}
 \left\{ \begin{array}{ccc}
 x^\prime & =& \frac{y}{\bar H}~,\\
 y^\prime &=& -3y-\frac{\bar{V}_x}{ \bar{H}}~,\end{array}\right.
 \end{eqnarray}
 where the prime represents the derivative with respect to $N$, $\bar{H}=\frac{H}{H_0}$   and $\bar{V}=\frac{V}{H_0^2M_{pl}^2}$. Moreover, the Friedmann equation now looks as  
 \begin{eqnarray}\label{friedmann}
 \bar{H}(N)=\frac{1}{\sqrt{3}}\sqrt{ \frac{y^2}{2}+\bar{V}(x)+ \bar{\rho}_{\chi}(N)+\bar{\rho}_{matt}(N) }~,
 \end{eqnarray}
where we have introduced the following dimensionless energy densities
 $\bar{\rho}_{\chi}=\frac{\rho_{\chi}}{H_0^2M_{pl}^2}$ and 
 $\bar{\rho}_{matt}=\frac{\rho_{matt}}{H_0^2M_{pl}^2}$.
Then, we have to integrate the dynamical system, starting at $N_{eq}=-8.121$, with initial conditions $x_{eq}$ and $y_{eq}$, and the value of the parameter $\tilde{M}$ is obtained equaling at $N=0$ the equation (\ref{friedmann}) to $1$, i.e., imposing $\bar{H}(0)=1$.

\

For the first case (the decay after the end of kination), the initial conditions 
are obtained analytically in  
equations (\ref{eq0}) and (\ref{doteq}). 
Effectively, from formula 
(\ref{teta})
in Subsection \ref{grav-heavy},
\begin{eqnarray}
y_{eq}\cong 2.82\times 10^{-35}\Theta^{-1}\frac{T_{rh}}{\mbox{GeV}}
\cong\frac{T_{rh}}{\mbox{GeV}}\left(\frac{m_X}{M_{pl}}  \right)^4
\left\{\begin{array}{cc}
   37  & \mbox{c.c.}  \\
  0.11   & \mbox{n.c.},
\end{array}
\right.
\end{eqnarray}
where c.c. means that the $\chi$-field is conformally coupled to gravity and n.c. non-conformally coupled.
Then,  for viable reheating temperatures $T_{rh}\leq 10^9$ GeV and as we will see in Subsection \ref{grav-heavy} for $m_{\chi}\cong 10^{15}$ GeV,  one has $y_{eq}\ll 1$.
And
for $x_{eq}$, after a simple calculation,
\begin{eqnarray}
x_{eq}\cong \sqrt{\frac{2}{3}}\left(1-\ln(2\sqrt{\Theta})\right)+\left(2-\frac{\pi}{6}\sqrt{\frac{g_*}{15}}      \right)\frac{T_R^2}{H_{kin}\Theta M_{pl}}, \qquad
y_{eq}\cong 1.7\frac{T_{eq}^3T_R}{H_0M_{pl}^2H_{kin}\Theta}.
\end{eqnarray}

\

Last, the initial conditions for the second case (the decay before the end of kination) are

\begin{eqnarray}
x_{eq}\cong \sqrt{\frac{2}{3}}\left(2+\ln\left(
\frac{3}{\pi}\sqrt{\frac{5}{g_*}\frac{H_{kin}M_{pl}}
{T_R^2}} \right)\right), \quad y_{eq}\cong \frac{4\pi}{9}\sqrt{\frac{g_{eq}}{5}}\left( \frac{g_{eq}}{g_*}\right)^{1/4}\frac{T_{eq}^3}{H_0M_{pl}T_R}.
\end{eqnarray}
 
\

\subsubsection{Numerical results}

In all cases compatible with the constraints found in this manuscript, which are summarized in Table \ref{table} (see Conclusions), we have obtained that $M/M_{pl}\cong 10^{-13}$, which coincides with the result obtained in \cite{pv}. 
Note that for this model, the energy scale of inflation $V^{1/4}(\varphi\ll -M_{pl})\sim \lambda^{1/4}M_{pl}\sim 10^{15}$ GeV is close to the GUT scale, while the energy scale for dark energy $V^{1/4}(\varphi\cong 0)\sim \lambda^{1/4}M\sim 10^{2}$ GeV is near the electroweak scale.

\

Next we show in Figures \ref{plotscaling1} and \ref{plotscaling2} the reduced densities $\{\bar{\rho}_i\}_{i=\chi,m,\varphi}$, the density parameters $\{\Omega_i\}_{i=\chi,m,\varphi}$ and the effective Equation  of State (EoS)  parameter $\omega_{eff}$ for all of them, showing that at 
the present time $w_{eff}\cong -0.6<-1/3$, which proves the current cosmic acceleration, and at 
late time $w_{eff}$ goes to $-1$, meaning that  this model leads to an eternal acceleration. As clearly seen in the figures, the results remain almost unchanged for the different considered cases, corresponding to different values of the reheating temperature, given that for all of them the value of parameter $\tilde{M}$ yields almost the same value, namely $\tilde{M}\sim 10^{13}$ GeV $\sim 10^{-5}M_{pl}$.

\begin{figure}[H]
    \centering
    \includegraphics[width=17cm]{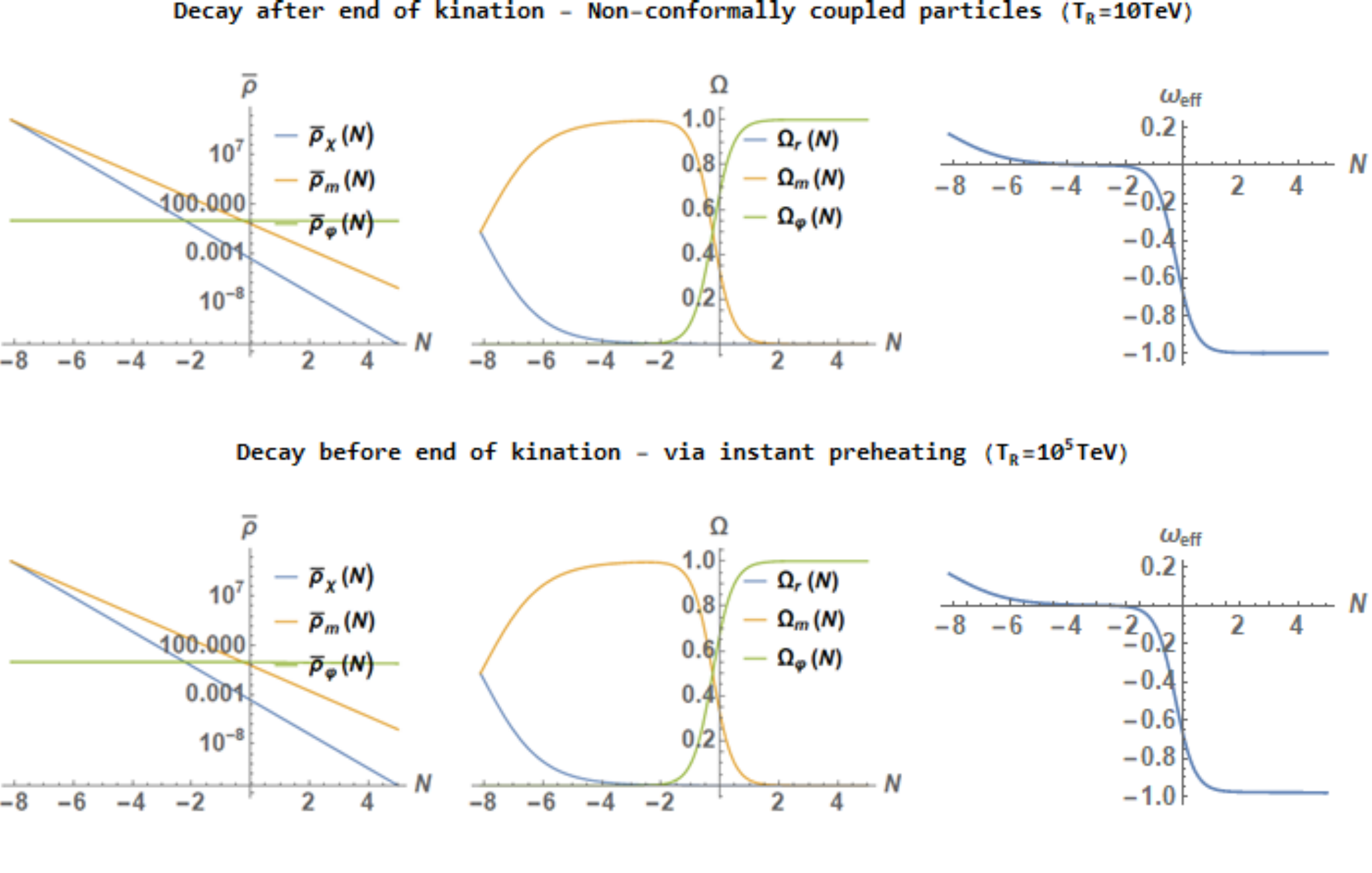}
    \caption{Numerical results for the allowed reheating mechanisms for the potential in \eqref{PV}.}
    \label{plotscaling1}
\end{figure}
\begin{figure}[ht]
    \centering
    \includegraphics[width=17cm]{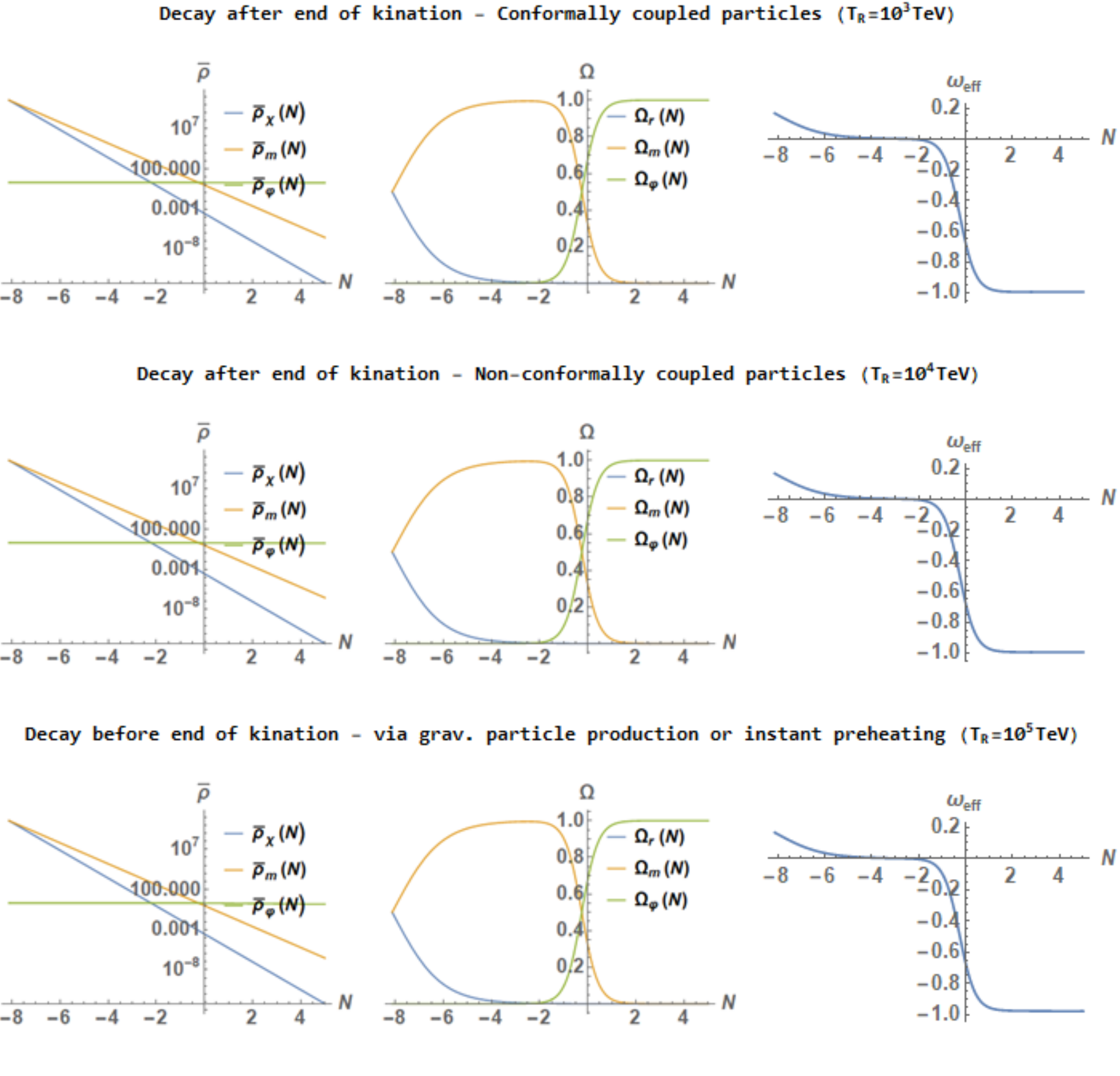}
    \caption{Numerical results for the allowed reheating mechanisms for the potential in \eqref{PV1}.}
    \label{plotscaling2}
\end{figure}

\

\subsection{Compatibility of the model with the cosmological perturbations}

After having studied the dynamics of the model, in order to verify its compatibility with the cosmological perturbations, we are going to compare the number of e-folds for our considered potentials, namely $N=\frac{2}{1-n_s}$, with the one obtained from \cite{liddle}
\begin{eqnarray}
\frac{k_*}{a_0H_0}=e^{-N}\frac{H_*}{H_0}\frac{a_{end}}{a_{kin}}\frac{a_{kin}}{a_R}\frac{a_R}{a_M}\frac{a_M}{a_0}=e^{-N}\frac{H_*}{H_0}\frac{a_{end}}{a_{kin}}\frac{\rho_R^{-1/12}\rho_M^{1/4}}{\rho_{kin}^{1/6}}\frac{a_M}{a_0},
\end{eqnarray}
where $M$ symbolizes the beginning of the matter domination era. Analogously as in \cite{AresteSalo:2017lkv}, it leads to
\begin{eqnarray}
N\cong 54.8+\ln\left(\frac{a_{end}}{a_E} \right)+\frac{1}{2}\ln\epsilon_*-\frac{1}{3}\ln\left(\frac{g_R^{1/4}T_RH_{kin}}{M_{pl}^2} \right),
\end{eqnarray}
where $g_R = 107$, $90$ and $11$ respectively for $T_R \geq 175$ GeV, $175 \mbox{ GeV } \geq T_R \geq 200$
MeV and $200 \mbox{ MeV } \geq T_R \geq 1$ MeV; $\ln\left(\frac{a_{end}}{a_E} \right)=\int_{H_{kin}}^{H_{end}}H(t)dt$, which has been numerically calculated for both considered potentials when we take the spectrum index to be the central value $n_s=0.968$, and $\epsilon_*=\frac{3}{16}(1-n_s)^2$. 

\

Therefore, we obtain the value of the reheating temperature in function of $n_s$ for both potentials, which has been represented in Figure \ref{boundsTR}. We observe that all the important bounds for our model, namely the BBN ones and the ones summarized in Table \ref{table} (see conclusions) lay within the allowed values for the spectral index, namely $n_s=0.968\pm 0.006$ \cite{Planck}.

\begin{figure}[H]
    \centering
    \includegraphics[width=17cm]{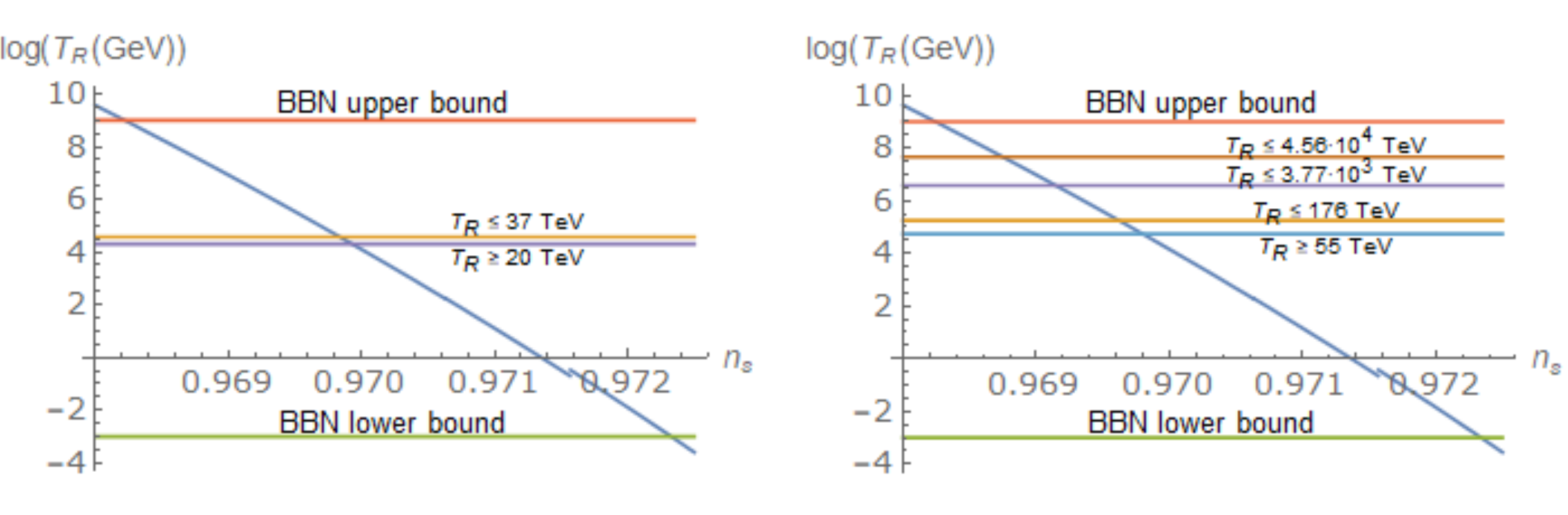}
    \caption{Relation between the reheating temperature $T_R$ in function of the spectral index $n_s$ for the potential in \eqref{PV} (left) and the one in \eqref{PV1} (right), with the corresponding bounds for $T_R$ found in this work.}
    \label{boundsTR}
\end{figure}

\

\section{Reheating in quintessential inflation}
\label{sec-IV}

 In this section we will discuss the most common ways to reheat the universe: 
 Reheating via gravitational production of light or superheavy particles and {\it instant preheating}.

\subsection{Gravitational production of light particles}\label{grav-light}

When the produced particles during the phase transition are very light, { the energy density of the relativistic plasma formed by these light particles is given by  \cite{pv,ford, Birrell, Haro, Giovannini1, Damour}   }
\begin{eqnarray}
 \rho_{\chi}(\tau)\cong R H_{kin}^4\left( \frac{a_{kin}}{a(\tau)} \right)^4, \end{eqnarray}
where $R\sim 10^{-2} N_s$, being $N_s$ the number of scalar fields, which for the minimal GUT is $4$ (the electro-weak Higgs doublet) \cite{pv}. So, we will use that 
$R\cong 10^{-1}$.

\begin{remark}
Here it is important to recall that this formula is only obtained for toy models (see for instance \cite{ford, Damour}) and we understand that it will also work for more realistic models in which the adiabatic evolution is broken near the beginning of the kination phase.
\end{remark}

Since as we immediately {show} the thermalization process of the plasma is an instantaneous process, the universe will become reheated at the end of the kination epoch, that is, when the energy densities of the scalar field and that of the relativistic plasma were of the same order. This occurs when
$ \left( \frac{a_{kin}}{a_{R}} \right)^2=\Theta$,
 where we have used, once again,  the so-called {\it heating efficiency} defined 
 in Subsection \ref{dyn-analytical} as
  $\Theta\equiv  \frac{\rho_{\chi, kin}}{\rho_{\varphi, kin}}$ and the fact that the energy density of the produced particles decays as $a^{-4}$ while the one of the inflation field decays as $a^{-6}$ during kination. Thus,  the reheating temperature is given by
 \begin{eqnarray}
 T_R=\left( \frac{30}{\pi^2 g_*} \right)^{1/4}\rho_{\chi, R}^{1/4}= { \left( \frac{3}{\pi^2 g_*} \right)^{1/4}\sqrt{\Theta}H_{kin}},
 \end{eqnarray}
 where $g_*=106.75$ are the degrees of freedom for the Standard Model. Then,
 \begin{eqnarray}
 T_R\cong { 3.33\times 10^{-7}\sqrt{\Theta}M_{pl}={ 8.12}\times 10^{11} \sqrt{\Theta}\mbox{ GeV}}
  \end{eqnarray}
and, since a simple calculation leads to the value $\Theta\cong \frac{H_{kin}^2}{30M_{pl}^2}\cong 6.9\times 10^{-14}$, we can conclude that the reheating temperature when the reheating is via the gravitational production of light particles is $T_R\cong { 213}$ TeV, which is basically the same result as the one obtained by { Peebles and Vilenkin} in their paper \cite{pv}.

\

Finally, we will show that 
the thermalization is nearly an instantaneous process compared with the duration of the kinetic era.
Following the reasoning of \cite{Spokoiny} and \cite{pv}, the decay products have a typical energy of the form  $\bar{\epsilon}\sim H_{kin}\left(\frac{a_{kin}}{a(\tau)}\right)$ and their number density is  $n\sim R\bar{\epsilon}^3\cong 10^{-1}\bar{\epsilon}^3$.
Now, we take into account that, if the particles interact
by the exchange of gauge bosons and establish thermal
equilibrium among the fermions and gauge bosons,
 the interaction rate will be $n\sigma$, where the cross section is given by $\sigma\sim \frac{\alpha^2}{\bar{\epsilon}^2}$, with the coupling constant satisfying the inequality $10^{-2}\leq \alpha\leq 10^{-1}$. Therefore, the thermal equilibrium will be accomplished when the interaction rate becomes comparable to the
Hubble parameter $H=H_{kin}\left(\frac{a_{kin}}{a(t)}\right)^3$, which happens when $\left(\frac{a_{kin}}{a_{th}}\right)^2=10^{-1}\alpha^2$, where the subscript ``${th}$'' attached to any quantity refers to its value at the time when the thermal equilibrium has been established. On the other hand, one can  calculate the scale factor at the reheating time $t = t_{R}$, which occurs at the end of kination, as follows:
Since $
 \left( \frac{a_{kin}}{a_{R}} \right)^2=\Theta$,
 then we will have
  \begin{eqnarray*}
 \left( \frac{a_{kin}}{a_{R}} \right)^2=\frac{\rho_{\chi, kin}}{\rho_{\varphi, kin}} = 10^{-1}\frac{H_{kin}^2}{3M_{pl}^2}=\frac{H_{kin}^2}{3M_{pl}^2\alpha^2}\left( \frac{a_{kin}}{a_{th}} \right)^2 
 \Longrightarrow a^2_{th}=\frac{H_{kin}^2}{3M_{pl}^2\alpha^2}a_R^2 \end{eqnarray*}
and, thus, 
\begin{eqnarray*}
a_{th}
\cong  \frac{8.3}{\alpha}\times 10^{-7}a_R\leq 8.3\times 10^{-5}a_R\Longrightarrow \frac{t_R}{t_{th}}\geq 1.7\times 10^{12} ,
\end{eqnarray*}
where we have used that during kination the scale factor evolves as $t^{1/3}$.
This result
 means that the thermal equilibrium occurs well before the equality between the energy density of the scalar field and the one of the decay products, i.e., well before to the end of kination which in this case coincides with the beginning of the radiation era. Hence,
one can safely assume an instantaneous thermalization.

\subsection{Gravitational production of superheavy particles}\label{grav-heavy}

In this subsection we will assume that the $\chi$-field has a mass $m_{\chi}$ greater than $10^{15}$ GeV. So, since this mass is greater than the Hubble rate,   we can apply the WKB solution when we
calculate the evolution of the modes.
In the appendix of \cite{deHaro:2017nui} it has been shown that the leading term of the re-normalized energy density of the produced particles
after the phase transition is given by 
\begin{eqnarray}
 \rho_{\chi}(\tau)=\frac{1}{2\pi^2 a^4(\tau)}\int_0^{\infty} \omega_k(\tau)k^2|\beta_k|^2dk
 \end{eqnarray}
and, for our model, {in order to obtain the $\beta$-Bogoliubov coefficient we use the WKB approximation.

Note that
the second iteration $W_k^{(2)}$ including temporal derivatives up to order four was obtained in  \cite{Bunch}, which is enough for our calculations because for the model (\ref{PV}) the third derivative of the Hubble rate is discontinuous and the term responsible for the leading contribution  to the $\beta$-Bogoliubov coefficient  is contained in $W_k^{(2)}$. As we show in Appendix \ref{appA},
 for the conformally coupled case, i.e. when $\xi=1/6$, the  term leading to the main contribution is given by $\frac{a^6m_{\chi}^2}{16\omega_k^5}\dddot{H}$, and for the non-conformally coupled case  by $-\frac{3a^4(\xi-1/6)}{4\omega_k^3}\dddot{H}$. Therefore, using equation (\ref{beta}) of Appendix \ref{appB} and the energy density of the produced particles
 $ \rho_{\chi}(\tau)\cong \frac{m_{\chi}}{2\pi^2a^3(\tau)}\int_0^{\infty} k^2 |\beta_k|^2 dk  $, we will obtain
\begin{eqnarray}\label{particleproduction}
\rho_{\chi}(\tau)\cong 
 \left\{ \begin{array}{cc}
 \frac{7\lambda^2}{589824\pi}\left( \frac{\dot{\varphi}_{kin}}{m_{\chi}}\right)^4\left( \frac{a_{kin}}{a(\tau)} \right)^3 & \mbox{ for the conformally coupled case}\\
 & \\
\frac{ \lambda^2}{{256}\pi}\left( \frac{\dot{\varphi}_{kin}}{m_{\chi}}\right)^4\left( \frac{a_{kin}}{a(\tau)} \right)^3 & \mbox{ for the nonconformally coupled case},
 \end{array}\right.
  \end{eqnarray}
where in the nonconformally coupled case we {have taken $\left|\xi-\frac{1}{6}\right|\cong 1$,
which is its maximum value because the WKB approximation is a perturbative one that only holds when $m_{\chi}^2\gg \left|\xi-\frac{1}{6}\right|R$ and, thus, since at GUT scales $R\cong 12H^2\cong 10^{29} \mbox{ GeV}^2$ and $M_{pl}\gg m_{\chi}\geq 10^{15}$ GeV, we can conclude that 
$\left|\xi-\frac{1}{6}\right|\leq 1$.}

 {
 
 \begin{remark}
 Note that in the nonconformally coupled case in \eqref{beta} we have corrected by a factor of $2$ the result obtained in \cite{haro19}. 
 \end{remark}
 
 }

As we will see in next section dealing with the overproduction of GW, when considering reheating via gravitational production of superheavy particles, in order to prevent the BBN success we have to impose the decay of the $\chi$-field to be after the end of kination, that is, after the equality between the energy density of the field and the one of the produced particles. 
 
 \
 
Since the decay is after  $t_{end}$ ($t_{end}$ denotes the instant when kination ends), one has to impose ${\Gamma}\leq H_{end}$, where $\Gamma$ is the decay rate of $\chi$-particles. Taking this into account, one has 
\begin{eqnarray}\label{31}
H^2_{end}=\frac{2\rho_{\varphi, end}}{3M_{pl}^2}\quad \mbox{and} \quad \rho_{\varphi, end}=\rho_{\varphi, kin}\left( \frac{a_{kin}}{a_{end}} \right)^6=3H^2_{kin}M_{pl}^2\Theta^2,
\end{eqnarray}
where we have used that for a superheavy field the {\it heating efficiency} satisfies $\Theta= \left( \frac{a_{kin}}{a_{end}} \right)^3$.
On the other hand, a simple calculation leads to the result
\begin{eqnarray}\label{teta}
\Theta
\cong \left\{\begin{array}{cc}
{ 7.61}\times 10^{-37}\left( \frac{M_{pl}}{m_{\chi}}\right)^4 & \mbox{ for the conformally coupled case} \\
&\\
{ 2.51}\times 10^{-34}\left( \frac{M_{pl}}{m_{\chi}}\right)^4 & \mbox{ for the nonconformally coupled case}.
\end{array}\right.
\end{eqnarray}

\

Consequently, from eqn. (\ref{31}) one can easily find $H_{end}=\sqrt{2}H_{kin}\Theta$ and, thus,
 one obtains that the decay rate has to 
 satisfy $\Gamma \leq \sqrt{2}H_{kin}\Theta$, which means that
 \begin{eqnarray}
  \Gamma \leq \left\{ \begin{array}{cc}
 3.7\times 10^{-24}\left( \frac{M_{pl}}{m_{\chi}}\right)^4 \mbox { GeV} & \quad \mbox{ for the conformally coupled case} \\
  &\\
  {1.25}\times 10^{-21}\left( \frac{M_{pl}}{m_{\chi}}\right)^4 \mbox { GeV} & \quad \mbox{ for the nonconformally coupled case}.  \end{array}\right. 
 \end{eqnarray}
 
  Since as we have already shown the thermalization  is nearly instantaneous, the reheating temperature (i.e., the temperature of the universe when the thermalized plasma starts to dominate) will be
\begin{eqnarray}
T_R=\left( \frac{30}{\pi^2 g_*} \right)^{1/4}\rho_{\chi, dec}^{1/4}= \left( \frac{90}{\pi^2 g_*} \right)^{1/4}\sqrt{{\Gamma}M_{pl}},
\end{eqnarray}
where we have used that after  $t_{end}$ the energy density of the produced particles dominates the energy density of the inflaton field. 
Then, we will have that $T_R\cong  { 0.54} \sqrt{\frac{\Gamma}{M_{pl}}} M_{pl}$ and, thus, we have the following bound for the reheating temperature,

\begin{eqnarray}
 T_R\leq  \left\{\begin{array}{cc}
{ 1.62}\times 10^{-3}\left( \frac{M_{pl}}{m_{\chi}}\right)^2 \mbox{GeV}& \quad \mbox{ for the conformally coupled case} \\
&\\
{ 2.99\times 10^{-2}}\left( \frac{M_{pl}}{m_{\chi}}\right)^2 \mbox{GeV}& \quad \mbox{ for the nonconformally coupled case} 
.\end{array}\right.
\end{eqnarray}

And finally, since we are assuming that the mass of the $\chi$-field is greater than $10^{15}$ GeV, we obtain the following upper bound for the reheating temperature,
\begin{eqnarray}
T_R\leq \left\{ \begin{array}{cc}
{ 9.64} \mbox{ TeV}& \quad \mbox{ for the conformally coupled case} \\
&\\
{178} \mbox{ TeV}& \quad \mbox{ for the  nonconformally coupled case} .
\end{array}\right.
\end{eqnarray}

\
 
 We end this subsection noting that these bounds are obtained without taking into account the production of GWs, which leads to more restrictive bounds as we will see in next section.

 \subsection{Instant preheating}\label{grav-instant}

{
In this subsection we will assume that the bare mass of the $\chi$-field, which we impose to be conformally coupled to gravity, is zero and 
 we also consider an
interaction between the inflaton field $\varphi$ and the quantum $\chi$-field, whose interacting Lagrangian is given by
 $ {\mathcal L}_{int}=-\frac{1}{2}g^2\varphi^2 \chi^2$, where $g$  is a dimensionless coupling constant.
 {The enhanced symmetry point has been chosen $\varphi=0$ because at this point the velocity of the scalar field is nearly maximum as one can see in Figure \ref{retrat}}.
  In this situation the $\chi$-particles, {which have an effective mass
 $m_{\chi, eff}(t)=g|\varphi(\tau)|$,}
  are created via a mechanism named {\it instant 
 preheating},  which was   introduced in \cite{fkl0}  in the framework of standard inflation and was applied for the first time to quintessential inflation in \cite{fkl}.
 
\

\begin{remark}
The reheating via {\it instant preheating} is usually used in models with very smooth potentials because in these models the gravitational production of particles is completely inefficient due to the adiabatic regime during all the evolution (see for instance \cite{dimopoulos, hossain2}). On the contrary, the introduction of the interacting Lagrangian term depicted 
above breaks down the adiabatic evolution at the beginning of the kination phase, which as we will see allows the production of enough particles to reheat the universe in a viable way. However, as we will see in Subsection \ref{dark-instant},
if we assume that dark matter is also created during the phase transition from inflation to kination, then this dark matter cannot be created via {\it instant preheating} and one needs another mechanism to create it, which could be the gravitational particle production. Therefore, in this hypothetical situation the potential cannot be so smooth. 
\end{remark}

 }

 \

  Then, if the reheating is via {\it instant preheating}, 
 soon after the beginning of kination the $\chi$-field acquires an effective mass equal to $m_{\chi, eff}=gM_{pl}$
 and the energy density of the $\chi$-field is given by \cite{fkl}
 \begin{eqnarray}
 \rho_{\chi}(\tau)=gM_{pl}n_{\chi}(\tau)=gM_{pl}n_{\chi, kin}\left(\frac{a_{kin}}{a(\tau)}   \right)^3, \end{eqnarray}
 where the number density of particles at the beginning of kination is { calculated as follows:
 
 \
 
 Near the beginning of kination, i.e. when $\varphi=0$, one has
 $\varphi(\tau)\cong \varphi'_{kin}(\tau-\tau_{kin})$ and the frequency of the $k$-mode of the field $\chi$ is 
 $\omega_k(\tau)=\sqrt{k^2+g^2a_{kin}^2\varphi'^2_{kin}(\tau-\tau_{kin})^2}$, where the expansion of the universe is not considered and for this reason we have approximated the scale factor by its value at the beginning of kination. Then, the $k$-mode of the $\chi$-field satisfies the equation of a time dependent harmonic oscillator 
 \begin{eqnarray}
 \chi_k''+\omega_k^2(\tau)\chi_k=0,
 \end{eqnarray}
 obtaining an over-barrier problem in scattering theory, whose $\beta$-Bogoliubov coefficient is related to the reflexion coefficient via the formula \cite{Popov, Marinov, Nikishov, Haro03}
 \begin{eqnarray}
   |\beta_k|^2=e^{-\mbox{Im}\left( \int_{\gamma}\omega_k(\tau)d\tau \right)},  
 \end{eqnarray}
 where $\gamma$ denotes a closed path that wraps around the  turning points $\tau_{\pm}=\tau_{kin}\pm i\frac{k}{ga_{kin}\varphi'_{kin}}$
 and the average number of produced particles in the $k$-mode is given by 
 \begin{eqnarray}
 n_k=|\beta_k|^2=e^{-\frac{\pi k^2}{ga_{kin}\varphi'_{kin}}}.
 \end{eqnarray}
 Thus, the average number density of $\chi$-particles at the beginning of kination is given by
  \begin{eqnarray}
 n_{\chi,kin}\equiv \frac{1}{2\pi^3a_{kin}^3}\int_0^{\infty}
 k^2n_kdk
 =\frac{g^{3/2}\dot{\varphi}_{kin}^{3/2}}{8\pi^3}.
 \end{eqnarray}
 
 }
 
 \begin{remark}
 Since the effective mass of the $\chi$-field is $g|\varphi(\tau)|$, in order to prevent the vacuum polarization effects from affecting the evolution of the inflation field during inflation, one has to impose the effective mass of the $\chi$-field to be greater than the Hubble rate, which leads to the condition
 \begin{eqnarray}
 g|\varphi(\tau)|\geq \sqrt{\frac{\lambda}{3}}M_{pl}\Longrightarrow g\geq \sqrt{\frac{\lambda}{3}}\frac{M_{pl}}{|\varphi(\tau)|},
 \end{eqnarray}
 { which always holds if we assume that $g\geq \sqrt{\frac{\lambda}{3}}\frac{M_{pl}}{|\varphi_{END}|}$ (because $|\varphi(\tau)|$ is a decreasing function during inflation)},
 where $\varphi_{END}$ denotes the value of the $\varphi$-field at the end of inflation. Since inflation ends when the slow-roll parameter $\epsilon$ is equal to one, one easily gets that $\varphi_{END}= \sqrt{\frac{3}{2}}\ln(\sqrt{3}(2-\sqrt{3}))M_{pl}\cong -0.94 M_{pl}$, which leads to the constraint for the parameter $g$ of
 \begin{eqnarray}
 g\geq 5.8 3 \times 10^{-6}.
 \end{eqnarray}

 \end{remark}

 These particles are very massive and, in order to avoid a second inflationary epoch due to the $\chi$-field, one has to assume that the decay is well before the end
 of the kination regime \cite{fkl}.
    Since the thermalization is nearly instantaneous as we have already seen, in this case the reheating is completed at the end of kination and, thus, the reheating temperature is calculated as follows:
 
 \
 
 Using that $ \frac{H_{dec}}{H_{kin}}=   \frac{\Gamma}{H_{kin}}=\left(\frac{a_{kin}}{a_{dec}} \right)^3$ we have
 \begin{eqnarray}
\rho_{\varphi,dec}=3{\Gamma}^2M_{pl}^2, \quad \mbox{and} \quad  \rho_{\chi, dec}=\frac{g^{5/2}M_{pl}\dot{\varphi}^{3/2}_{kin}}{8\pi^3}\frac{\Gamma}{H_{kin}}\cong
1.85\times 10^{-5} g^{5/2} M_{pl}^3\Gamma.
\end{eqnarray} 
 On the other hand, from the condition  $\rho_{\chi, dec} \leq \rho_{\varphi, dec}$ (the decay is before the end of the kination phase), one gets
\begin{eqnarray}\label{xx}
\Gamma\geq 6.18\times 10^{-6} g^{5/2}M_{pl}
\end{eqnarray}
and, since we have shown that $g\geq 5.8{ 3}\times 10^{-6}$, we see that the decay rate is greater than $5.0{7}\times 10^{-19} M_{pl}\cong 1.2{}$ GeV. We note that the evolution of the energy density of the created particles and the background are respectively
\begin{eqnarray}
\rho_{\chi}(t)=\rho_{\chi, dec}\left( \frac{a_{dec}}{a(t)} \right)^4 ,\qquad 
\rho_{\varphi}(t)=\rho_{\varphi, dec}\left( \frac{a_{dec}}{a(t)} \right)^6, 
\end{eqnarray}
which tells us that at the time when  
the kination phase ends, i.e., when 
$\rho_{\varphi, end}=\rho_{\chi, end}$, 
one has
$ 
\left( \frac{a_{dec}}{a_{end}} \right)^2=\frac{\rho_{\chi, dec}}{\rho_{\varphi,dec}}$. So, the reheating temperature takes the form 
\begin{eqnarray}\label{reheating1}
 T_R=  \left(\frac{30}{\pi^2 g_*} \right)^{1/4}  \rho_{\chi, end}^{1/4}
  =\left(\frac{30}{\pi^2 g_*} \right)^{1/4}\rho_{\chi, dec}^{1/4}\sqrt{\frac{\rho_{\chi, dec}}{\rho_{\varphi,dec}}} 
\nonumber \\ \cong {6.7}\times 10^{-5} g^{15/8}\left(\frac{M_{pl}}{\Gamma}  \right)^{1/4} M_{pl}\cong {1.63}\times 10^{14}g^{15/8}\left(\frac{M_{pl}}{\Gamma}  \right)^{1/4}
\mbox{ GeV}.
\end{eqnarray}
This reheating temperature [i.e., eqn. (\ref{reheating1})] could be bounded using (\ref{xx}), obtaining
\begin{eqnarray}\label{reheatingtemperature}
 T_R\leq { 3.27}\times 10^{15} g^{5/4}\mbox{ GeV}.
\end{eqnarray}
On the other hand,  since the decay is after the beginning of kination, we have that $\Gamma\leq H_{kin}\cong 1.44\times 10^{-6} M_{pl}\cong 3.51\times 10^{12}$ GeV, getting the bound
\begin{eqnarray}
T_R\geq { 4.7}\times 10^{15}g^{15/8}
\mbox{ GeV},
\end{eqnarray}
which means that, in order to 
preserve the BBN success, a reheating temperature approximately between $1$ MeV and  $10^6$ TeV is required and, thus, 
 the constraint 
$g\leq 
{ 2.76}\times 10^{-4} $ has to be satisfied, which restricts the value of $g$ in the following narrow band,
\begin{eqnarray}
5.8{ 3}\times 10^{-6} \leq  g\leq { 2.76}\times 10^{-4}{ \Longrightarrow g\cong 10^{-5}}.
  \end{eqnarray}
Finally, if for example we choose
$g\cong 10^{-5}$ and $\Gamma\cong 10^{-10} M_{pl},$
which satisfy the constraint (\ref{xx}),  one obtains a reheating temperature equal to
\begin{eqnarray}
T_R
\cong { 2.18}\times 10^{4} \mbox{ TeV}.
\end{eqnarray}

\

\section{BBN constraints  coming from the production of  Gravitational Waves}
\label{sec-BBN}

This section is devoted to present the bounds of the proposed improved version of the  {\it quintessential inflationary} model using the Big Bang  Nucleosynthesis (BBN), where we explicitly use the BBN constraints from the logarithmic spectrum of GWs and consequently the BBN bounds from the overproduction of GWs.

\subsection{BBN constraints from the logarithmic spectrum of GWs}
\label{subsec-gw1}

It is well-known that during inflation GWs  are produced (known as primordial GWs, in short PGWs) and in the post-inflationary period, i.e., during kination, the logarithmic spectrum of GWs, namely
$\Omega_{GW}$ defined as $\Omega_{GW}\equiv \frac{1}{\rho_c}\frac{d\rho_{GW}(k)}{d\ln k }$ (where $\rho_{GW}(k)$ is the energy density spectrum of the produced GWs; $\rho_c=3H_0^2M_{pl}^2$, where $H_0$ is the present value of the Hubble parameter, is the so-called {\it critical density}) scales as $k^2$ \cite{rubio}, producing a spike in the spectrum of GWs at high frequencies. Then, so that GWs do not destabilize the BBN, the following bound must be imposed  (see Section 7.1 of \cite{maggiore}),
\begin{eqnarray}\label{integral}
I\equiv h_0^2\int_{k_{BBN}}^{k_{end}} \Omega_{GW}(k) d \ln k \leq 10^{-5},
\end{eqnarray}
where $h_0\cong 0.678$ parametrizes the experimental uncertainty to determine the current value of the Hubble constant and $k_{BBN}$, $k_{end}$ are the momenta associated to the horizon scale at the BBN and at the end of inflation respectively. As has been shown in \cite{Giovannini1}, the main contribution of the integral \eqref{integral} comes from the modes that leave the Hubble radius before the inflationary epoch and finally re-enter during the kination, that means, for $k_{end}\leq k\leq k_{kin}$, where
$k_{end}=a_{end}H_{end}$ and $k_{kin}=a_{kin}H_{kin}$.  For these modes one can calculate the  logarithmic spectrum of GWs as in \cite{Giovannini} (see also \cite{rubio, Giovannini2, Giovannini3,Giovannini:2016vkr} where the graviton spectra
in quintessential models have been reassessed, in a model-independent way, using numerical techniques),

\begin{eqnarray}\label{Omega}
\Omega_{GW}(k)=\tilde{\epsilon}\Omega_{\gamma}h^2_{GW} \left(\frac{k}{k_{end}}  \right)\ln^2\left(\frac{k}{k_{kin}}  \right),
\end{eqnarray}
where $h^2_{GW}=\frac{1}{8\pi}\left(\frac{H_{kin}}{M_{pl}}  \right)^2$
is the amplitude of the GWs; $\Omega_{\gamma}\cong 2.6\times 10^{-5} h_0^{-2}$ is the present density fraction of radiation, and the quantity $\tilde{\epsilon}$, which is approximately equal to $0.05$ for the Standard Model of particle physics,  takes into account the variation of massless degrees of freedom between decoupling and thermalization (see \cite{rubio, Giovannini1} for more details). As has been derived in \cite{Giovannini1}, the specific form of the expression above comes from the behavior of the Hankel functions for small arguments. Now, plugging expression (\ref{Omega}) into (\ref{integral}) and disregarding the sub-leading logarithmic terms, one finds  
\begin{eqnarray}\label{constraintx}
 2\tilde{\epsilon}h_0^2\Omega_{\gamma}h^2_{GW}\left( \frac{k_{kin}}{k_{end}} \right)\leq 10^{-5}
 \Longrightarrow 
 10^{-2}\left( \frac{H_{kin}}{M_{pl}} \right)^2\left( \frac{k_{kin}}{k_{end}} \right)\leq 1.   \end{eqnarray}
  
  \

  \begin{remark}
 A further bound on primordial gravitational waves is imposed by the CMB  constraint on additional massless degrees of freedom.  As GWs with frequencies larger than the
 corresponding horizon at CMB decoupling  contribute to the radiation density of the Universe, constraints on their total energy density can be phrased in terms of the effective number of massless neutrino species $N_{eff}$, which is bounded by $N_{eff}=3.04\pm 0.17$  (see Section $5$ of \cite{pieroni}), namely
 \begin{eqnarray}
 \int_{k_{BBN}}^{k_{end}} \Omega_{GW}(k) d \ln k =1.95\times 10^{-5}(N_{eff}-3.046)\Longrightarrow \nonumber\\
 \Longrightarrow 
 \int_{k_{BBN}}^{k_{end}} \Omega_{GW}(k) d \ln k\leq 0.6513\times 10^{-5} \ \ \  \mbox{ at $2\sigma$ C.L.}
  \end{eqnarray}
 and, thus,   turning to the following constraint,
 \begin{eqnarray}
  3.33\times10^{-2}\left( \frac{H_{kin}}{M_{pl}} \right)^2\left( \frac{k_{kin}}{k_{end}} \right)\leq 1,  \end{eqnarray} 
  \end{remark}
which is practically the same constraint obtained above. {Note that this constraint is more restrictive than the one obtained with the effective number of massless neutrino species from BBN, namely $N_{eff}=3.28\pm 0.28$ \cite{pieroni}, which leads to the same constraint as in Equation \eqref{constraintx}.

 \

To calculate the ratio $k_{kin}/k_{end} $,  we  will have to study the following
 three different situations:
\begin{enumerate}
\item When the produced particles are very light and its energy density decays as $a^{-4}$.
In this case, as has been shown in  \cite{rubio},  one will have 
\begin{eqnarray}
\frac{k_{kin}}{k_{end}} =\frac{1}{\sqrt{2} \Theta},
\end{eqnarray}
where $\Theta$ is once again the {\it heating efficiency} introduced previously. Thus,  the constraint (\ref{constraintx}) eventually leads to 
\begin{eqnarray}
\Theta  \geq 7\times 10^{-3} \left(\frac{H_{kin}}{M_{pl}}  \right)^2\cong 1.45\times 10^{-14}.\end{eqnarray}

Taking into account that when the reheating is due to the creation of very light particles during the phase transition the reheating temperature is (see Subsection \ref{grav-light})
\begin{eqnarray}
T_R={ \left(\frac{3}{\pi^2 g_*} \right)^{1/4}\sqrt{\Theta} H_{kin}\cong 8.{ 12}\times 10^{11} \sqrt{\Theta} \mbox{ GeV}},
\end{eqnarray}
one has the following lower bound,
\begin{eqnarray}
T_R\geq { 97.8}  \mbox{ TeV}.
\end{eqnarray}

Finally, note that we have shown that when reheating is due to the gravitational production of light particles the reheating temperature is $T_R\cong {213}$ TeV, which means that the reheating via the gravitational production of light particles satisfies the bound (\ref{integral}).

\item When the reheating is due to the production of superheavy particles which decay after the end of kination, as we have shown in Subsection \ref{grav-heavy}, we have that
$\Theta=\left(\frac{a_{kin}}{a_{end}} \right)^3$ and $H_{end}=\sqrt{2}H_{kin}\Theta$ and, thus,
\begin{eqnarray}
\frac{k_{kin}}{k_{end}}=\frac{a_{kin}H_{kin}}{a_{end}H_{end}}=\frac{\Theta^{1/3}}{\sqrt{2}\Theta}=\frac{1}{\sqrt{2}\Theta^{2/3}}.
\end{eqnarray}
So, the constraint (\ref{constraintx}) leads to
\begin{eqnarray}\label{2b}
\Theta^{2/3}  \geq 7\times 10^{-3} \left(\frac{H_{kin}}{M_{pl}}  \right)^2\cong 1.4{5}\times 10^{-14}\Longrightarrow \Theta\geq 1.7{5}\times 10^{-21}.\end{eqnarray}

Now, since in Subsection \ref{grav-heavy} we obtained the following value of the {\it heating efficiency},
\begin{eqnarray}
\Theta
\cong \left\{\begin{array}{cc}
7.{61}\times 10^{-37}\left( \frac{M_{pl}}{m_{\chi}}\right)^4 & \mbox{ for the conformally coupled case} \\
&\\
{2.51}\times 10^{-34}\left( \frac{M_{pl}}{m_{\chi}}\right)^4 & \mbox{ for the nonconformally coupled case},
\end{array}\right.
\end{eqnarray}
we deduce that in the conformally coupled case the mass of the $\chi$-field has to satisfy $m_{\chi}\leq 3.5{2}\times 10^{14}$ GeV,
which is incompatible with our assumption $m_{\chi}\geq 10^{15}$ GeV.
This shows that the gravitational production of superheavy particles conformally coupled to gravity is not viable.
On the contrary,
when the $\chi$-field is not
conformally coupled to gravity, one gets the bound $m_{\chi}\leq 1.{5}\times 10^{15}$ GeV, which means that the viability of our model requires its mass to be $m_{\chi}\cong 10^{15}$ GeV when reheating is due to the gravitational production of superheavy particles noncoformally coupled to gravity.

\item When the decay happens before the end of kination, as the case of {\it instant reheating},
a simple calculation leads to 
\begin{eqnarray}
\frac{k_{kin}}{k_{end}} =\frac{1}{\sqrt{2} \Theta}\left( \frac{{\Gamma}}{H_{kin}}  \right)^{1/3}
\end{eqnarray}
and, consequently, the constraint (\ref{constraintx}) becomes
\begin{eqnarray}\label{49}
\Theta \left(\frac{H_{kin}}{{\Gamma}}  \right)^{1/3}\geq 7\times 10^{-3} \left(\frac{H_{kin}}{M_{pl}}  \right)^2\Longrightarrow 
\left(\frac{M_{pl}}{{\Gamma}}  \right)^{1/4}\geq 2.4{2}\times 10^{-2} \left(\frac{H_{kin}}{M_{pl}\Theta^{3/5}}  \right)^{5/4},
\end{eqnarray}
which applied to our model finally leads to another lower bound of the reheating temperature, which is obtained via {\it instant preheating} {(see Subsection \ref{grav-instant})},
\begin{eqnarray}\label{40}
T_R\geq { 1.97}\times 10^5\frac{g^{15/8}}{\Theta^{3/4}} \mbox{ GeV}\cong 1.{84\times 10^6 }g^{3/8} \mbox{ GeV},
\end{eqnarray}
where we have used that in the case of {\it instant preheating} one has $\Theta=\frac{g^2}{2\pi^2}$. 

Now, since the reheating temperature has to be less than $10^6$ TeV, one gets the bound  $g\leq {1.97\times 10^7}$, which is less restrictive than the one obtained in Subsection \ref{grav-instant}, meaning that, when reheating is due to the production of particles via {\it instant preheating}, the bound (\ref{integral}) is clearly overpassed.

\end{enumerate}

\subsection{BBN bounds from the overproduction of GWs}
\label{subsec-gw2}

The success of the BBN demands that \cite{dimopoulos}
\begin{eqnarray}\label{bbnconstraint}
\frac{\rho_{GW,R}}{\rho_{\chi, R}}\leq 10^{-2},
\end{eqnarray}
where $\rho_{GW}(t)$ is the energy density of the GWs produced  at the phase transition and both quantities are evaluated at the reheating time.  The value of the energy density of the GWs is 
 $\rho_{GW}(t)\cong 10^{-2} H^4_{kin} \left( \frac{a_{kin}}{a(t)} \right)^4$ (see for example \cite{pv, Damour}).

 \

 Then, when the reheating is via the gravitational production of light particles, we have
 \begin{eqnarray}
\frac{\rho_{GW,R}}{\rho_{\chi, R}}= \frac{\rho_{GW,kin}}{\rho_{\chi, kin}}\cong \frac{1}{N_s},
\end{eqnarray} 
which, as pointed out by Peebles and Vilenkin, results $\frac{1}{N_s}=0.25$ for a minimal GUT. So, this bound is never reached, meaning that in this case the overproduction of GWs could affect the BBN process. However, if one goes beyond a minimal GUT and  accepts minimal supersymmetric (SUSY) theories, then $N_s=104$ and, thus, the bound (\ref{bbnconstraint}) is overpassed.

 \

On the other hand, in the case in which superheavy particles (which could decay in lighter ones to match with the HBB) are gravitationally created during the phase transition, we firstly see that the decay can never be before the end of kination because, if so, at the decay time, which occurs when $H_{dec}=\Gamma$,  we would have
\begin{eqnarray}\label{45}
\frac{\rho_{GW,dec}}{\rho_{\chi, dec}}= \frac{\rho_{GW,kin}}{\rho_{\chi, kin}}\left(\frac{\Gamma}{H_{kin}} \right)^{1/3}=
\frac{10^{-2}}{{3}}\left( \frac{H_{kin}}{M_{pl}} \right)^2\frac{1}{\Theta}\left(\frac{\Gamma}{H_{kin}} \right)^{1/3}\cong
\nonumber\\ \cong { 6.91 \times 10^{-15}}\frac{1}{\Theta}\left(\frac{\Gamma}{H_{kin}} \right)^{1/3},
\end{eqnarray}
where, once again, we have used that $\left(\frac{a_{kin}}{a_{dec}}\right)^3=\frac{\Gamma}{H_{kin}}$. Thus, for the nonconformally coupled case (recall that the conformally coupled case was disregarded by the bound (\ref{integral})),  we obtain 
\begin{eqnarray}
\frac{\rho_{GW,R}}{\rho_{\chi, R}}=\frac{\rho_{GW,dec}}{\rho_{\chi, dec}}=  { 2.75}\times 10^{19}\left( \frac{m_{\chi}}{M_{pl}}\right)^4\left(\frac{\Gamma}{H_{kin}} \right)^{1/3}\cong {2.44}\times 10^{21}\left( \frac{m_{\chi}}{M_{pl}}\right)^4\left(\frac{\Gamma}{M_{pl}} \right)^{1/3}
\end{eqnarray}
and now we use that the decay is before the end of kination, i.e., that $\rho_{\chi,dec}\leq \rho_{\varphi,dec}$.
{ Taking into account  that
$
\rho_{\chi,dec}=\frac{\lambda^2}{{256}\pi}\left(
\frac{\dot{\varphi}_{kin}}{m_{\chi}}
\right)^4\left( \frac{a_{kin}}{a_{dec}} \right)^3$ and
$
\left( \frac{a_{kin}}{a_{dec}} \right)^3=\frac{\Gamma}{H_{kin}}$},
one gets  the bound 
\begin{eqnarray}
\left(\frac{\Gamma}{M_{pl}} \right)^{1/3}\geq {7.12}\times 10^{-14}\left( \frac{M_{pl}}{m_{\chi}}\right)^{4/3}\end{eqnarray}
and, thus,
\begin{eqnarray}
\frac{\rho_{GW,R}}{\rho_{\chi, R}}\geq {1.74}\times 10^{8}\left( \frac{m_{\chi}}{M_{pl}}\right)^{8/3}.\end{eqnarray}
So, imposing the constraint (\ref{bbnconstraint}), one gets that
$m_{\chi}\leq { 3.5}\times 10^{14}$ GeV, which contradicts our assumption that $m_{\chi}\geq 10^{15}$ GeV.

\

 Hence, the decay must occur after $t_{end}$ and, assuming once again the instantaneous thermalization,  the reheating time will coincide with the decay one. Then, we will have 
$\rho_{\chi, dec}=3{\Gamma}^2M_{pl}^2$ and, since 
{\begin{eqnarray}
H_{dec}=H_{end}\left( \frac{a_{end}}{a_{dec}} \right)^{3/2}\Longrightarrow \left( \frac{a_{end}}{a_{dec}} \right)^{3/2}=\frac{\Gamma}{\sqrt{2}H_{kin}\Theta},
\end{eqnarray}
we obtain
\begin{eqnarray}
\rho_{GW,dec}=\rho_{GW,end}\left( \frac{a_{end}}{a_{dec}} \right)^4= \rho_{GW,end}\left( \frac{\Gamma}{\sqrt{2}H_{kin}\Theta}    \right)^{8/3}
=10^{-2} H^4_{kin}\Theta^{-4/3}\left( \frac{\Gamma}{\sqrt{2}H_{kin} }  \right)^{8/3}\end{eqnarray}}
and, thus, 
\begin{eqnarray}
\frac{\rho_{GW,R}}{\rho_{\chi,R}}\cong \frac{\rho_{GW,dec}}{3\Gamma^2M_{pl}^2}
\cong 
10^{-2} \frac{H^4_{kin}}{3\Gamma^2M_{pl}^2}\Theta^{-4/3}\left( \frac{\Gamma}{\sqrt{2}H_{kin} }  \right)^{8/3}
\cong 2.15 \times 10^{-11}\Theta^{-4/3}\left( \frac{\Gamma}{M_{pl} }  \right)^{2/3}.
\end{eqnarray}
Now, we use that for the nonconformally coupled case we have already shown that $m_{\chi}\cong 10^{15}$ GeV and
$\Theta={ 2.51}\times 10^{-34}\left(\frac{M_{pl}}{m_{\chi}}  \right)^4\cong { 8.90\times 10^{-21}}$ in order to get that
\begin{eqnarray}
\frac{\rho_{GW,R}}{\rho_{\chi,R}}\cong { 1.17\times 10^{16}}\left( \frac{\Gamma}{M_{pl} }  \right)^{2/3}
\Longrightarrow \frac{\Gamma}{M_{pl}}\leq { 7.90\times 10^{-28}},
\end{eqnarray}
which reduces the maximum reheating temperature $T_R\cong { 0.54}\sqrt{\frac{\Gamma}{M_{pl}}}M_{pl}$ to be $T_R\leq { 37}$ TeV.

Finally, in the case of {\it instant preheating}, when the decay is before the end of kination as we have already explained, using the formula
(\ref{45}) and the fact that $\Theta=\frac{g^2}{2\pi^2}$ we arrive at
\begin{eqnarray}
\frac{\rho_{GW,R}}{\rho_{\chi,R}}\cong
{ 1.21}\times 10^{-11}\frac{1}{g^2}\left( \frac{\Gamma}{M_{pl}} \right)^{1/3}
\end{eqnarray}
and, taking into account the bound $ \frac{\Gamma}{M_{pl}}\leq 1.44\times 10^{-6}$, we get
\begin{eqnarray}
\frac{\rho_{GW,R}}{\rho_{\chi,R}}\leq
{ 1.3{6}}\times 10^{-13}\frac{1}{g^2}.
\end{eqnarray}

{ 

\

So, 
since at the end of Subsection \ref{grav-instant} we have already shown that $g\geq 5.{8}3\times 10^{-6}$, we reach 
\begin{eqnarray}
\frac{\rho_{GW,R}}{\rho_{\chi,R}}\leq
{ 1.3{ 6}}\times 10^{-13}\frac{1}{g^2}
\leq 4.{02}\times 10^{-3},
\end{eqnarray}
 which assures that the constraint 
\eqref{bbnconstraint} is fulfilled when the reheating is via {\it instant preheating}. }

 \
 
 { Summing up, we have shown that a viable reheating in the case of gravitational reheating requires the creation of superheavy nonconformally coupled particles with  mass nearly $10^{15}$ GeV, which must decay after the end of kination, obtaining a  maximum reheating temperature around { $37$} TeV. And, when the reheating is via {\it instant preheating}, the coupling constant $g$ has to be close to $10^{-5}$, obtaining a reheating temperature greater than
{ $ 20$} TeV (where we have used  the bound  (\ref{40})). }
   
\

\section{Abundance of dark matter}

{ In this section we will explore the possibility that 
the breakdown of the adiabatic regime
 leads to the possibility to explain the abundance of dark matter through the gravitational production of superheavy particles \cite{hashiba, hashiba1}, although gravitational production of dark matter could also occur in standard inflation during the oscillations of the inflaton field \cite{kolb1, ema,kolb2, kolb3} {{}(see also the earlier papers
\cite{starobinsky, Starobinsky1, Vilenkin}). }
}

\subsection{Reheating via production of superheavy nonconformally coupled to gravity}\label{dark-heavy}

{
As we have already shown, the reheating via gravitational production of superheavy particles conformally coupled is not viable and, when these superheavy particles are nonconformally coupled to gravity, its mass must be very close to $10^{15}$ GeV.

\

Now we also assume that there is another kind of  superheavy particles conformally coupled to gravity, named $Y$-particles,  which do not decay  and only interact gravitationally, and could be the responsible for the current abundance of the dark matter. 
In Subsection \ref{grav-heavy} we have seen that the energy density of the $Y$-particles will be
$
\rho_{Y}(\tau)\cong 
 \frac{7\lambda^2}{589824\pi}\left( \frac{\dot{\varphi}_{kin}}{m_{Y}}\right)^4\left( \frac{a_{kin}}{a(\tau)} \right)^3 $, where $m_Y$ is the mass of the $Y$-particles,  and the one of the $\chi$-particles evolves before the decay as  
$
\rho_{\chi}(\tau)\cong 
\frac{ \lambda^2}{{256}\pi}\left( \frac{\dot{\varphi}_{kin}}{m_{\chi}}\right)^4\left( \frac{a_{kin}}{a(\tau)} \right)^3$.

\

Therefore, since in order to preserve the BBN success the decay of the $\chi$-particles has to be after the end of kination, as we have seen in Section 3.1, the thermalization is instantaneous. So, at the reheating time we will have
\begin{eqnarray}
\frac{\rho_{Y,R}}{\rho_{\chi,R}}=\frac{7}{{2304}}
\left(\frac{m_{\chi}}{m_Y}\right)^4.
\end{eqnarray}

Now, taking into account that $m_{\chi}\cong 10^{15}$ GeV and the energy density of the $\chi$-particles must be greater than the one of the $Y$-particles so that the universe reheats, we deduce the following bound for the mass of $Y$-particles,
\begin{eqnarray}
m_Y>\left(7/{2304} \right)^{1/4}m_{\chi}\cong {0.235} m_{\chi}.
\end{eqnarray}

After reheating the evolution of the corresponding energy densities will be
\begin{eqnarray}
\rho_{\chi}(\tau)=\rho_{\chi,R}\left(\frac{a_{R}}{a(\tau)} \right)^4\quad \mbox{ and} \quad   \rho_Y(\tau)=\rho_{Y,R}\left(\frac{a_{R}}{a(\tau)} \right)^3,\end{eqnarray}
meaning that at the matter-radiation equality 
\begin{eqnarray}
\frac{\rho_{\chi,eq}}{\rho_{Y,eq}}=\frac{{2304}}{7}
\left(\frac{m_{Y}}{m_{\chi}}\right)^4\frac{a_R}{a_{eq}}.
\end{eqnarray}

\

On the other hand, taking the following observational data at present time ($H_0\cong 1.42\times 10^{-33}$ eV, $\Omega_{Y,0}= 0.262$,  $\Omega_{b,0}=0.048$ and $\Omega_{matt,0}=0.31$, where
$b$ denotes the baryonic matter and $matt$ the total matter (dark+baryonic)), it is satisfied that
\begin{eqnarray}\label{eq}
\rho_{matt, eq}=\rho_{matt,0}(1+z_{eq})^3 \quad 
\rho_{Y, eq}=\rho_{Y,0}(1+z_{eq})^3, 
\end{eqnarray}
where $z_{eq}$ denotes once again the cosmic red-shift at the matter-radiation equality. Then, since $\rho_{\chi,eq}=
\rho_{matt,eq}$
 at the matter-radiation equality,
  we will have
$\rho_{\chi,eq}=\rho_{matt,0}(1+z_{eq})^3$ and, thus,
\begin{eqnarray}
\frac{\rho_{\chi,eq}}{\rho_{Y,eq}}=
\frac{\rho_{matt,0}}{\rho_{Y,0}}=
\frac{\Omega_{matt,0}}{\Omega_{Y,0}},
\end{eqnarray}
which, combined with (\ref{eq}), leads to the relation
\begin{eqnarray}
\frac{a_{R}}{a_{eq}}=\frac{\Omega_{matt,0}}{\Omega_{Y,0}}
\frac{7}{{2304}}\left(\frac{m_{\chi}}{m_{Y}}\right)^4\cong 
{3.6}\times 10^{-3}\left( \frac{m_{\chi}}{m_Y} \right)^4
\end{eqnarray}
and, consequently,
\begin{eqnarray}\label{eq1}
\rho_{Y,eq}\cong 
{4.67\times 10^{-8}}\rho_{Y,R}^{ren}\left( \frac{m_{\chi}}{m_Y} \right)^{12}
\cong {1.42\times 10^{-10}}\rho_{\chi,R}^{ren}\left( \frac{m_{\chi}}{m_Y} \right)^{16}\cong 
\nonumber\\
\frac{{1.42}\pi^2 g_*}{30}\times {10^{-10}}  T_{R}^4 \left( \frac{m_{\chi}}{m_Y} \right)^{16}\cong
{ 3.15\times 10^{-63}} T_{R}^4 \left( \frac{M_{pl}}{m_Y} \right)^{16},\end{eqnarray}
where we have used that $m_{\chi}\cong 10^{15} \mbox{ GeV}.$

\

Now, taking into account that 
$\rho_{Y,eq}=3H_0^2M_{pl}^2\Omega_{Y,0}(1+z_{eq})^3$ and
choosing $3365$ as the value of the cosmic red-shift at matter-radiation equality, we obtain
$\rho_{Y,eq}\cong 3.5{98}\times 10^{-1} \mbox{ eV}^4$
and, inserting this expression in (\ref{eq1}), one gets the following relation between the reheating temperature and the mass of dark matter,
\begin{eqnarray}
T_R\cong \left( \frac{3.{598}}{{315}}\right)^{1/4}10^{16}\left( \frac{m_{Y}}{M_{pl}} \right)^{4}\mbox{ eV}\cong { 3.27} \times 10^6
\left(\frac{m_{Y}}{M_{pl}} \right)^{4}\mbox{ GeV}.
\end{eqnarray}

Then, since the decay is after the end of kination and the thermalization is nearly instantaneous, the reheating temperature is 
\begin{eqnarray}\label{temperature2}
T_{R}=
\left( \frac{90}{\pi^2 g_*} \right)^{\frac{1}{4}}\sqrt{{\Gamma}M_{pl}}
\cong 1.32 \times 10^{18} \sqrt{\frac{\Gamma}{M_{pl}}} \mbox{ GeV},
\end{eqnarray}
which gives us the following relation between the mass of dark matter and the decay rate of the $\chi$-particles,
\begin{eqnarray}
m_Y\cong {7.97}\times 10^2\left( \frac{\Gamma}{M_{pl}} \right)^{1/8}M_{pl}\cong 1.{94} \times 10^{21}\left( \frac{\Gamma}{M_{pl}} \right)^{1/8}\mbox{ GeV}.
\end{eqnarray}

\

Finally, we have to use the bounds that must satisfy the decay rate to bound the mass of dark matter. For example the overproduction  of gravitational waves leads to $\frac{\Gamma}{M_{pl}}\leq {7.90\times 10^{-28}}$ and, thus,
$m_Y\leq { 7.94}\times 10^{17}$ GeV.
And, taking into account that the reheating temperature must be greater than $1$ MeV because the BBN occurs approximately at $1$ MeV when the universe is already reheated, { from equation \eqref{temperature2}} one gets 
\begin{eqnarray}
\frac{\Gamma}{M_{pl}}\geq 5.74\times 10^{-43}\Longrightarrow
m_Y\geq {1.02 \times 10^{16}} \mbox{ GeV},
\end{eqnarray}
that is, a viable quintessential inflation model where dark matter is created gravitationally requires the mass of dark matter to be bounded as follows,
\begin{eqnarray}
{1.02\times 10^{16}} \mbox{ GeV}\leq m_Y 
\leq {7.94}\times 10^{17}  \mbox{ GeV},
\end{eqnarray}
and a maximum reheating temperature around { $37$} TeV.

}

\subsection{Reheating via {\it instant preheating}}\label{dark-instant}

{ 

First of all it is important to note that, when reheating is via {\it instant preheating}, the production of dark matter cannot be via the same mechanism because the coupling constant $g$ is restricted to be close to $10^{-5}$ and, thus, the energy of the dark matter and the one of the particles that reheat the universe after its decay would be of the same order, which would  forbid a radiation phase, which is essential for correctly depicting the evolution of our universe.

\

Therefore, in that case we also have to consider the possibility that dark matter was created gravitationally. 
So, we consider once again superheavy $Y$-particles only conformally coupled to gravity, which would be the responsible for the present abundance of dark matter in the universe in our model. 

}

\

{
As we have already  discussed, in order to avoid a second inflationary period  it is mandatory that,   unlike the superheavy particles created gravitationally studied in the previous section,
these $\chi$-particles   decay well before the end of kination. Then, at the matter-radiation equality we will have
\begin{eqnarray}
\rho_{\chi,eq}=\rho_{\chi,dec}\left( \frac{a_{dec}}{a_{eq}}  \right)^4, \quad \rho_{Y,eq}=\rho_{Y,dec}\left( \frac{a_{dec}}{a_{eq}}  \right)^3,\end{eqnarray}
and, thus,
\begin{eqnarray}
\frac{\rho_{\chi,eq}}{\rho_{Y,eq}}=
\frac{\rho_{\chi,dec}}{\rho_{Y,dec}}\left(\frac{a_{dec}}{a_{eq}}
\right)={
\frac{gM_{pl}n_{\chi,kin}}{\rho_{Y,kin}}\left(\frac{a_{dec}}{a_{eq}}
\right),}
\end{eqnarray}
{where we have  used  that at the decay time
$\rho_{\chi,dec}=gM_{pl}n_{\chi,kin}\frac{\Gamma}{H_{kin}}$
and $\rho_{Y,dec}=\rho_{Y,kin}\frac{\Gamma}{H_{kin}}$.
}

\

On the other hand, as we have already seen in the previous subsection,
\begin{eqnarray}
\frac{\rho_{\chi,eq}}{\rho_{Y,eq}}=
\frac{\Omega_{matt,0}}{\Omega_{Y,0}},
\end{eqnarray}
meaning that
\begin{eqnarray}
\frac{a_{dec}}{a_{eq}}=\frac{\Omega_{matt,0}}{\Omega_{Y,0}}{
\frac{\rho_{Y,kin}}{gM_{pl}n_{\chi,kin}}.}
\end{eqnarray}
Therefore, we will have
\begin{eqnarray}
\rho_{Y,eq}=\rho_{\chi,eq}\frac{\Omega_{Y,0}}{\Omega_{matt,0}}
=
{
gM_{pl}n_{\chi,kin}\frac{\Gamma}{H_{kin}}
\left(\frac{\Omega_{matt,0}}{\Omega_{Y,0}}\right)^3
\left(\frac{\rho_{Y,kin}}{gM_{pl}n_{\chi,kin}}\right)^4}
\end{eqnarray}
and, recalling  that 
\begin{eqnarray}
{\frac{\rho_{Y,kin}}{gM_{pl}n_{\chi,kin}}=
\frac{7\pi^2\lambda^2}{2048}6^{-3/4}g^{-{5/2}}
\frac{H_{kin}^{5/2}M_{pl}^{3/2}}{m_Y^4}}
\quad \mbox{ and }\quad 
n_{\chi,kin}=\frac{{6^{3/4}}(gH_{kin}M_{pl})^{3/2}}{8\pi^3},
\end{eqnarray}
we get that
\begin{equation}
  \rho_{Y,eq}=     {{1.08\times 10^{-42}}g^{-15/2} }\left(\frac{M_{pl}}{m_Y} \right)^{16}\frac{\Gamma}{M_{pl}}\mbox{ eV}^4,  
\end{equation}
but the energy density of the dark matter at the matter-radiation equality is $\rho_{Y,eq}\cong 3.5{98}\times 10^{-1} \mbox{ eV}^4$. Then, we have the following relation between the mass of the dark matter and the decay rate of the $\chi$-particles
\begin{eqnarray}
m_Y\cong {{2.54\times 10^{-3}} g^{-{1}5/32}}
\left( \frac{\Gamma}{M_{pl}}\right)^{1/16} M_{pl}\cong {{5.61\times 10^{-1}}
\left( \frac{\Gamma}{M_{pl}}\right)^{1/16} M_{pl}},
\end{eqnarray}
where we have used that $g\cong 10^{-5}$.

\

Finally, using that when the reheating is via {\it instant preheating}  the decay of the $\chi$-particles will be during the kination phase, we have the bounds obtained in Subsection \ref{grav-instant},
\begin{eqnarray}
6.18\times 10^{-6} g^{5/2}\cong 1.95\times 10^{-18}\leq \frac{\Gamma}{M_{pl}}\leq  1.44\times 10^{-6},
\end{eqnarray}
which bound the mass of the dark matter to be in the domain
\begin{eqnarray}
{ { 1.07\times 10^{17}} \mbox{ GeV}}\leq m_Y\leq { { 5.90\times 10^{17}} \mbox{ GeV}}\Longrightarrow m_Y\cong 10^{17} \mbox{ GeV},
\end{eqnarray}
that is, the mass of dark matter must be very close to $10^{17}$ GeV when reheating is via {\it instant preheating}.

}

\

{
\section{Other kind of potentials}
In this section we would like to check  the importance of the breakdown of the adiabatic evolution. For this reason we will consider a more abrupt phase transition than the one given by the potential (\ref{PV}). For example, we choose the following potential,
\begin{eqnarray}\label{PV1}
V(\varphi)=\left\{\begin{array}{ccc}
\lambda M_{pl}^4\left(1-e^{\alpha\frac{\varphi}{M_{pl}}}\right) + \lambda M^4 & \mbox{for} & \varphi\leq 0\\
\lambda\frac{M^8}{\varphi^{4}+M^4} &\mbox{for} & \varphi\geq 0,\end{array}
\right.
\end{eqnarray}
where now $\alpha$ denotes a positive dimensionless parameter. Note that, in this case, the inflationary piece is an Exponential SUSY inflation-type potential.

\

Here, when the field vanishes the potential has a discontinuity in its first derivative,
which was pointed out in \cite{starobinsky0,starobinsky1}, and could be obtained introducing a second scalar field experiencing a first order phase transition. This means, using Raychaudhuri equation, that the second derivative of the field and, thus, the second derivative of the Hubble rate are discontinuous at the beginning of kination. So, once again, we have to understand this model as a toy model, which allows us to perform analytically all the calculations, belonging to the class of potentials satisfying 
$\frac{1}{\omega_k^4(\tau)}\frac{d^3\omega_k(\tau)}{d\tau^3}\geq 1$ near the beginning of kination, where $\omega_k(\tau)=\sqrt{k^2+m_{\chi}^2a^2(\tau)}$ is the frequency of the $k$-mode of the $\chi$-field.
Another important thing is that, since the phase transition is abrupter than in the previous case, this means that now the gravitational production will be greater, thus obtaining a greater reheating temperature than for the potential (\ref{PV}).

}

\

{

As proved in \cite{hap19}, both potentials lead to equivalent expressions for the spectrum index, number of e-folds and ratio of tensor to scalar perturbations, {that is, the spectral index is 
$n_s\cong 1-\frac{2}{N}$ and the tensor/scalar ratio  is given by $r=\frac{8}{\alpha^2N^2}$, where the number of
e-folds depends on the reheating temperature as follows,
\begin{eqnarray}
N\cong 60-\frac{1}{3}\ln\left( \frac{T_R}{\mbox{GeV}} \right).
\end{eqnarray}

Hence,} the computations will be done for the same value of $\alpha$ as used in the previous potential, namely $\alpha=\sqrt{\frac{2}{3}}$, for which we obtain the corresponding numerical value of $\dot{\varphi}_{kin}=5.75\times 10^{-6} M_{pl}^2$.

\

If we first consider the case in which the reheating takes place via the gravitational production of light particles,
{ the energy density of produced particles is, once again, given by $\rho_{\chi}(\tau)\cong 10^{-1}
H_{kin}^4\left( \frac{a_{kin}}{a(\tau)} \right)^4$ and, following step by step the reasoning done in Subsection 
\ref{grav-light}, we get that} the reheating temperature becomes greater than before, namely $T_R\cong 568$ TeV. 

\

With regards to the case of gravitational production of superheavy particles, here we first need to recalculate the $\beta$-Bogoliubov coefficients both when the $\chi$ field is conformally coupled to gravity or not. {In this case we only need the first order WKB approximation, i.e. $W_k^{(1)}$, which is obtained in Appendix \ref{appA}, and the value of the $\beta$-Bogoliubov coefficient (see formula (\ref{beta1}) of Appendix \ref{appB}), to obtain the energy density of the produced particles at the beginning of kination
\begin{eqnarray}
\rho_{\chi,kin}=\left\{\begin{array}{ll} { \frac{5\lambda^2\dot{\varphi}_{kin}^2}
{196608}\left(\frac{M_{pl}}{m_{\chi}}  \right)^2} & \mbox{for $\xi=1/6$} \\
&\\
 \frac{3\lambda^2\dot{\varphi}_{kin}^2}
{256\pi}\left(\frac{M_{pl}}{m_{\chi}}  \right)^2  & \mbox{for $\left|\xi-\frac{1}{6} \right|\cong 1$.}\end{array}\right.
\end{eqnarray}

\

}Hence, analogously as done before, we arrive at the value of the heating efficiency, namely
\begin{eqnarray}
\Theta\cong\left\{\begin{array}{ll} 4.12\times 10^{-25}\left(\frac{M_{pl}}{m_{\chi}} \right)^2 & \mbox{for $\xi=1/6$} \\ 
&\\
{ 6.04}\times 10^{-23}\left(\frac{M_{pl}}{m_{\chi}} \right)^2 & \mbox{for $\left|\xi-\frac{1}{6} \right|\cong 1$,}  \end{array} \right.
\end{eqnarray}
and the lower bound for the decay rate, {when the decay is after the end of kination,} is
\begin{eqnarray}
\Gamma\leq\left\{\begin{array}{ll} 3.34\times 10^{-12}\left(\frac{M_{pl}}{m_{\chi}} \right)^2 & \mbox{for $\xi=1/6$} \\
&\\
 { 4.89}\times 10^{-10}\left(\frac{M_{pl}}{m_{\chi}} \right)^2 & \mbox{for $\left|\xi-\frac{1}{6} \right|\cong 1$,}  \end{array} \right.
\end{eqnarray}
which leads {for the decay being after the end of kination} to a reheating temperature of
\begin{eqnarray} \label{trpot2}
T_R\leq \left\{\begin{array}{ll} 1.54\times 10^3\frac{M_{pl}}{m_{\chi}}\text{GeV} &  \mbox{for $\xi=1/6$,} \\ 
&\\
1.{87}\times 10^4\frac{M_{pl}}{m_{\chi}}\text{GeV} & \mbox{for $\left|\xi-\frac{1}{6}\right|\cong 1$,} \end{array}\right.
\end{eqnarray}
which is respectively $3.77\times 10^3$ TeV and ${4.56}\times 10^4$ TeV when restricting $m_{\chi}\geq 10^{15}$GeV, { so we obtain minimum reheating temperatures considerably greater than the ones obtained for the potential with a discontinuity in the second derivative}. Now, when considering instant preheating, the results are very similar to the other potential, obtaining as well a very narrow band for $g$ corresponding to $g\cong 10^{-5}$.

\

Then, by taking into account the BBN constraints, we first consider the ones coming from the logarithmic spectrum of GW. When the reheating is produced via the production of light particles, we get the constraint $T_R\geq 260$ TeV, which is fulfilled by the computed value $T_R\cong 568$ TeV. With regards to the production of superheavy particles decaying after the end of kination we obtain the bounds $m_{\chi}\leq 1.80\times 10^{16}$ GeV for the conformally coupled case and $m_{\chi}\leq {2.17}\times 10^{17}$ GeV when non-conformally coupled. Note that, differently from the other potential, both constraints are compatible with $m_{\chi}\geq 10^{15}$ GeV. 

\

{

As we will immediately  see, when dealing with the overproduction of GWs, for this potential the decay of {superheavy particles} is possible before the end of kination. In this case, the
reheating temperature is given by (\ref{reheating1}), namely
\begin{eqnarray}
 T_R=  \left(\frac{30}{\pi^2 g_*} \right)^{1/4}  \rho_{\chi, end}^{1/4}
  =\left(\frac{30}{\pi^2 g_*} \right)^{1/4}\rho_{\chi, dec}^{1/4}\sqrt{\frac{\rho_{\chi, dec}}
  {\rho_{\varphi,dec}}}, \end{eqnarray}
with 
\begin{eqnarray}
\rho_{\chi, dec}=\rho_{\chi, kin}\frac{\Gamma}{H_{kin}}
\quad \mbox{and}\quad \rho_{\varphi, dec}=3\Gamma^2M_{pl}^2,
\end{eqnarray}
which leads to the following reheating temperature.
\begin{eqnarray}
T_R\cong \left\{\begin{array}{ll} {4.07\times 10^{-5}\left(\frac{M_{pl}}{\Gamma} \right)^{1/4}
\left(\frac{M_{pl}}{m_{\chi}}  \right)^{3/2} \mbox{ GeV}} & \mbox{for $\xi=1/6$}\\ 
&\\
1.72\times 10^{-3}\left(\frac{M_{pl}}{\Gamma} \right)^{1/4}
\left(\frac{M_{pl}}{m_{\chi}}  \right)^{3/2} \mbox{ GeV} & \mbox{for $\left|\xi-\frac{1}{6} \right|\cong 1$,}\end{array}\right..
\end{eqnarray}

Now, applying the bound (\ref{49}), namely
\begin{eqnarray}
\left(\frac{M_{pl}}{{\Gamma}}  \right)^{1/4}\geq 2.4{2}\times 10^{-2} \left(\frac{H_{kin}}{M_{pl}\Theta^{3/5}}  \right)^{5/4}
\end{eqnarray}
with 
\begin{eqnarray}\Theta\cong \left\{ \begin{array}{ll} { 4.12\times 10^{-25}\left( \frac{M_{pl}}{m_{\chi}}\right)^2} \mbox{for $\xi=1/6$} \\ 
&\\
6.04\times 10^{-23}\left( \frac{M_{pl}}{m_{\chi}}\right)^2\mbox{for $\left|\xi-\frac{1}{6} \right|\cong 1$,}\end{array}  \right.
\end{eqnarray} 
the term 
$\left(\frac{M_{pl}}{m_{\chi}}  \right)^{3/2} $ appearing in the expressions  of $T_R$ and $\Theta$ cancels and
we get the following lower bound for the reheating temperature,
\begin{eqnarray}
T_R\geq \left\{\begin{array}{ll} { 176  \mbox{ TeV}} & \mbox{for $\xi=1/6$} \\ 176 \mbox{ TeV} & \mbox{for $\left|\xi-\frac{1}{6} \right|\cong 1$.}\end{array}\right.
\end{eqnarray}

}

\

As for the case of instant preheating. we get a {minimum} reheating temperature around $55$ TeV. {Thus, for this kind of potentials, if the particles decay before the end of kination the reheating via gravitational production of  superheavy particles could lead to a reheating temperature greater than the one obtained when the reheating is via {\it instant preheating}.

}

\

Regarding the constraints from the overproduction of GW,
{ reheating via production of light particles is also forbidden and},
in the case of the production of superheavy particles decaying { before} the end of kination, we obtain that
\begin{eqnarray} \label{102}
\left(\frac{\Gamma}{M_{pl}}\right)^{1/3}\geq \left\{ \begin{array}{ll} 9.89\times 10^{-10}\left(\frac{M_{pl}}{m_{\chi}}\right)^{2/3} & \mbox{for $\xi=1/6$} \\ 
&\\
5.21\times 10^{-10}\left(\frac{M_{pl}}{m_{\chi}}\right)^{2/3} & \mbox{for $\left|\xi-\frac{1}{6} \right|\cong 1$}, \end{array} \right.
\end{eqnarray}
which restricts $m_{\chi}$ to be $m_{\chi}\leq {9.94}\times 10^{14}$ GeV and $m_{\chi}\leq {1.2\times 10^{16}}$ GeV for the conformal and non-conformal cases respectively. Hence, both cases are compatible with the minimum mass for $m_{\chi}$. {And when superheavy particles decay after the end of kination, the upper bound for the reheating temperature is
\begin{eqnarray}
T_R\leq \left\{ \begin{array}{ll} 6.26\times 10^3 \text{ TeV} & \mbox{for $\xi=1/6$} \\ 9.20\times 10^5 \text{ TeV} & \mbox{for $\left|\xi-\frac{1}{6} \right|\cong 1$} \end{array} \right.,
\end{eqnarray}
which does not reduce the bounds in \eqref{trpot2} for $m_{\chi}\geq 10^{15}$ GeV.
}

\

{ Therefore,  the constraints  coming 
from the production of GWs lead to completely different results depending on how abrupt the phase transition is. For instance, for the first potential reheating via gravitational production of superheavy conformally coupled particles is forbidden, which does not happen with the second potential if these particles decay after the end of kination. In addition, for the second potential {the decay of superheavy particles, both conformally and non-conformally coupled to gravity,} could be produced before or after the end of the reheating, obtaining a very efficient reheating mechanism which could lead to reheating temperatures greater than the one obtained using {\it instant preheating} as a reheating mechanism. 

}

\

Dealing with the present abundance of dark matter,
we are going to consider two types of particles: $\chi$-particles, {which can be now both conformally and non-conformally coupled to gravity}, and Y-particles, conformally coupled and responsible for the abundance of  dark matter. In contrast with the former potential, the decay of $\chi$-particles can be both before and after the end of kination. If we first proceed analogously as in Subsection \ref{dark-heavy}, i.e. by considering that the decay is produced after the end of kination when $\chi$-particles are non-conformally coupled, we obtain that
\begin{eqnarray}
\rho_{Y,eq}\cong 1.{25\times 10^{-7}}T_R^4\left(\frac{m_{\chi}}{m_Y} \right)^8,
\end{eqnarray}
reaching finally the following bounds,
\begin{eqnarray}
{ 156}\leq \frac{m_Y}{m_{\chi}} \leq { 1.5}\times 10^8,
\end{eqnarray}
which result in the following range of values for the dark matter mass, namely ${1.56\times 10^{17}}\text{ GeV}\leq m_Y\leq 2.44\times 10^{18}$ GeV, taking into account that $10^{15}\text{ GeV}\leq m_{\chi}\leq {2.17}\times 10^{17}$ GeV and that $m_Y\leq M_{pl}$. {When $\chi$-particles are conformally coupled the obtained bounds are $1.84\times 10^{18}\leq m_Y\leq 2.44\times 10^{18}$, hence $m_Y\lesssim M_{pl}$}.

\

On the other hand, if the decay is produced before the end of kination and considering first $\chi$-particles non-conformally coupled, we have that
\begin{eqnarray}
\rho_{Y,eq}= \rho_{Y,dec}\left(\frac{a_{dec}}{a_{eq}}\right)^3\cong \rho_{Y,dec}(8.07\times 10^{-3})^3\left(\frac{m_{\chi}}{m_Y} \right)^6\cong \nonumber \\ \cong
\rho_{\chi,dec}\frac{\Omega_{Y,0}}{\Omega_{matt,0}}(8.07\times 10^{-3})^4\left(\frac{m_{\chi}}{m_Y} \right)^8 = \nonumber \\  =\frac{3\lambda^2\dot{\varphi}_{kin}^2\Omega_{Y,0}\Gamma}{256\pi\Omega_{matt,0}H_{kin}}\left(\frac{M_{pl}}{m_{\chi}} \right)^2(8.07\times 10^{-3})^4\left(\frac{m_{\chi}}{m_Y}\right)^8,
\end{eqnarray}
where we have used that $\frac{a_{dec}}{a_{eq}}=\frac{\Omega_{matt,0}}{\Omega_{Y,0}}\frac{\rho_{Y,dec}}{\rho_{\chi,dec}}\cong 8.07\times 10^{-3}\left(\frac{m_{\chi}}{m_Y} \right)^2$, given that $\rho_{a,dec}=\rho_{a,kin}\frac{\Gamma}{H_{kin}}$ for $a=\chi,Y$. Now, using equation \eqref{49} and equation \eqref{102}, we find that bounds for the dark matter mass are considerably over the Planck's mass. {The same happens when $\chi$-particles are conformally coupled.} Hence, the presence of dark matter gravitationally created during the phase transition from the end of inflation to the beginning of kination cannot be explained with {a reheating via the gravitational creation  of} superheavy particles decaying before the end of kination for this potential.

\

Finally, when the reheating is produced via instant preheating, the energy density of the dark matter particles at the matter-radiation equality results
\begin{eqnarray}
\rho_{Y,eq}\cong 3.14\times 10^5 g^{-15/2}\left(\frac{M_{pl}}{m_Y}\right)^8\frac{\Gamma}{M_{pl}} \text{eV}^4,
\end{eqnarray}
leading, as well, to bounds for $m_Y$ over the Planck's mass. Therefore, instant preheating cannot either be used for this potential in order to explain the presence of {gravitationally produced }dark matter.

}

\

\section{Concluding remarks}
\label{sec-conclu}

{
In the present work, we have studied with all details the reheating of the universe via gravitational particle production and via {\it instant preheating} in quintessential inflation, taking into account the bound imposed by the production of GWs during the phase transition between the end of inflation and the beginning of kination.

\

In order to perform analytically all the calculations, we have considered a toy model inspired in the well-known Peebles-Vilenkin model { in which the discontinuity occurs in the second derivative of the potential.}

\

Our study shows that the reheating via gravitational production of light particles is forbidden due to the overproduction of GWs, i.e., the bounds imposed to prevent the success of the BBN are not overpassed. A similar situation occurs when the reheating is via the gravitational production of superheavy particles conformally coupled to gravity, in this case the bound imposed by the spectrum of the GWs is not accomplished.
So, only two situations are acceptable to have a viable reheating that does not affect the success of the BBN:
\begin{enumerate}
    \item Reheating via graviational particle production of superheavy particles not conformally coupled to gravity.
    \item Reheating via {\it instant preheating}.
\end{enumerate}

\

However, several restrictions must be imposed to the parameters appearing in the theory: In the case of gravitational production of superheavy particles nonconformally coupled to gravity, the decay of theses particles in lighter ones in order to obtain a relativistic plasma has to be after the end of kination obtaining a maximum reheating temperature around {$37$} TeV. In addition, the mass of these superheavy particles has to be approximately equal to  $10^{15}$ GeV. 

\

On the contrary, when reheating is via {\it instant preheating}, the produced particles have to decay before the end of the kination phase, obtaining a minimum temperature around $10$ TeV. Moreover,  the dimensionless coupling constant between the inflaton field and these particles has to be  of the order of $10^{-5}$.

\

On the other hand, when one assumes that dark matter could be created via gravitational particle production of conformally coupled particles during the phase transition from the end of inflation to the beginning of kination,  its mass has to range between 
$10^{16}$ GeV and $10^{18}$ GeV, when the reheating is via gravitational production of superheavy particles nonconformally coupled to gravity. And when the reheating is via {\it instant preheating} the mass of the dark matter would only be {of the order of $10^{17}$ GeV}.

\

In last section, we have considered another toy model inspired in the Peebles-Vilenkin model in which the discontinuity occurs in the first derivative of the potential and we have shown the differences with respect to the first potential considered, i.e., with the one with the discontinuity in the second derivative of the potential. 
Basically, in that case, if the reheating is via gravitational production of superheavy particles the reheating temperature is considerably increased being able to be even greater than the one obtained when the reheating mechanism is via the {\it instant preheating}.

\

Moreover, the constraints coming from the production of GWs
allow in this case the decay of superheavy particles before and after the end of kination. However, if one assumes that the abundance of dark matter is due to its gravitational production during the phase transition, then neither the reheating via gravitational production of superheavy particles decaying before the end of kination nor via {\it instant preheating} could be able to explain this abundance.

\

Finally, we show the allowed cases with the corresponding values of the parameters in the following table, where c.c. and c.c. stand for $\chi$-particles being respectively conformally and non-conformally coupled to gravity, $V_1(\varphi)$ and $V_2(\varphi)$ refer to the two potentials that have been considered, and a line has been drawn where we have achieved no constraints because the corresponding process has been proved as forbidden.

\

\begin{table}[ht]
\begin{center} 
\begin{tabular}{ c  c | c | c | c | c | c | c |} 
\cline{3-8}
& & \multicolumn{6}{|c|}{REHEATING VIA} \\
\cline{3-8}
  & & \multicolumn{5}{|c|}{gravitational production of} &  \\  \cline{3-7}
 & & light & \multicolumn{4}{|c|}{superheavy particles decaying} & {\it instant} \\
 \cline{4-7}
 & & particles & \multicolumn{2}{c|}{after kination ends} & \multicolumn{2}{c|}{before kination ends} & {\it preheating} \\
 \cline{4-7}
 & &  & c.c. & n.c. &  $\ \ \ \ $ c.c $\ \ \ \ $  & n.c. & $g\cong 10^{-5}$ \\
 \hline
 \multicolumn{1}{|c|}{$V_1(\varphi)$} & $T_R$ (TeV) & 
 $\line(1,0){25}$
 & $\line(1,0){25}$
 & 
 $\lesssim 37$ 
 & $\line(1,0){25}$ & $\line(1,0){25}$
 &   {$\gtrsim 20$} \\

\cline{2-8}
\multicolumn{1}{|c|}{} & $m_{\chi}$ (GeV) &  & $\line(1,0){25}$ & $\cong 10^{15}$   & $\line(1,0){25}$ & $\line(1,0){25}$  &  \\
\cline{2-8}
\multicolumn{1}{|c|}{} & $m_Y$ (GeV) & $\line(1,0){25}$ & $\line(1,0){25}$ & $10^{16}-10^{18}$     & $\line(1,0){25}$& $\line(1,0){25}$ & $\cong 10^{17}$ \\ 
 \hline
 \multicolumn{1}{|c|}{$V_2(\varphi)$} & $T_R$ (TeV) & $\line(1,0){25}$  & $\lesssim 3.8\times 10^3$ & $\lesssim 4.6\times 10^4$ & $\gtrsim 180$ & $\gtrsim 180$  & $\gtrsim 55$\\
 
\cline{2-8}
\multicolumn{1}{|c|}{} & $m_{\chi}$ (GeV) &  & $10^{15}-10^{16}$ & $10^{15}-10^{17}$   & $\cong 10^{15}$ & $10^{15}-10^{16}$&  \\
\cline{2-8}
\multicolumn{1}{|c|}{} & $m_Y$ (GeV) & $\line(1,0){25}$ & { $\cong 10^{18}$} & $10^{17}-10^{18}$   & $\line(1,0){25}$ & $\line(1,0){25}$ & $\line(1,0){25}$  \\
 \hline
\end{tabular}
\end{center}
\caption{Constraints for $T_R$, $m_{\chi}$ and $m_Y$ for the different potentials and reheating mechanisms.} \label{table}
\end{table}

}

\section*{ACKNOWLEDGMENTS}
This investigation  has been supported by MINECO (Spain) grant  MTM2017-84214-C2-1-P, and  also in part by the Catalan Government 2017-SGR-247.

\

\begin{appendices}
\titlecontents{section}
  [1em]{}{\contentslabel{1em}\appendixname}{\hspace*{-1em}}
  {\titlerule*[1pc]{}\contentspage}
\titleformat{\section}
  {\normalfont\bfseries}{\appendixname \ \thesection.}{0.5em}{}

\section{The WKB approximation in cosmology} \label{appA}

For a non-conformally coupled with gravity $\chi$-field, its
 $k$-mode satisfies the equation \cite{Birrell}
\begin{eqnarray}
\chi_k''+\Omega_k^2\chi_k=0,
\end{eqnarray}
where $\Omega^2_k=\omega^2_k+\left(\xi-\frac{1}{6}\right)a^2R$, being $\xi$ the coupling constant to gravity, $R=6(\dot{H}+2H^2)$ the Ricci scalar and 
$\omega_k=\sqrt{k^2+a^2m_{\chi}^2}$ the time dependent frequency of the $k$-mode.
The solution of this equation for a positive frequency mode is
$\chi_k=\frac{1}{\sqrt{2W_k}}e^{-i\int^{\tau}W_k(\eta)d\eta}$, where $W_k$ satisfies the equation \cite{Bunch}
\begin{eqnarray}
W_k^2=\Omega_k^2-\frac{1}{2}\left(\frac{W_k''}{W_k}-\frac{3}{2}\frac{W_k'^2}{W_k^2}  \right).
\end{eqnarray}
Then, the WKB solution is obtained solving iteratively this equation, taking as a zero-order WKB solution $W_k^{(0)}=\Omega_k$.

\

The first iteration $W_k^{(1)}$ leads to
\begin{eqnarray}\label{first}
W_k^{(1)}\cong \omega_k+\frac{1}{2\omega_k}(\xi-1/6)a^2R
-\frac{1}{4\omega_k}\left(\frac{\omega_k''}{\omega_k}-
\frac{3}{2}\frac{(\omega_k')^2}{\omega_k^2}
\right),
\end{eqnarray}
and
the second iteration $W_k^{(2)}$ including temporal derivatives up to order four was obtained in  \cite{Bunch}.
The result of this calculation in terms of the cosmic time is given by 
\begin{eqnarray}\label{W}
W_{k}^{(2)}=\omega_k+\frac{(\xi-1/6)a^2}{\omega_k}(4H^2+3\dot{H})-\frac{m_{\chi}^2a^4}{4\omega_k^3}(\dot{H}+3H^2)+\frac{5m_{\chi}^2a^6}{8\omega_k^5}H^2
\nonumber\\
+\frac{m_{\chi}^2a^6}{16\omega_k^5}(\dddot{H}+15\ddot{H}H+10\dot{H}^2+86\dot{H}H^2+60H^4)
\nonumber \\
-\frac{m_{\chi}^4a^8}{32\omega_k^7}(28\ddot{H}H+19\dot{H}^2+394\dot{H}H^2+507H^4)
\nonumber \\
+\frac{221m_{\chi}^6a^{10}}{32\omega_k^9}(\dot{H}+3H^2)H^2-\frac{1105m_{\chi}^8a^{12}}{128\omega_k^{11}}H^4
\nonumber\\
-\frac{(\xi-1/6)a^4}{4\omega^3_k}(117H^4+198H^2\dot{H}+54\dot{H}^2+27\ddot{H}H+3\dddot{H})
\nonumber\\
+(\xi-1/6)\frac{m_{\chi}^2a^6}{8\omega_k^5}
(24H^4+87H^2\dot{H}+3\ddot{H}H+18\dot{H}^2).
\end{eqnarray}

\

 As a consequence, for the conformally coupled case, i.e. when $\xi=1/6$, the  term leading to the main contribution is given by $\frac{a^6m_{\chi}^2}{16\omega_k^5}\dddot{H}$, and for the nonconformally coupled case  by $-\frac{3a^4(\xi-1/6)}{4\omega_k^3}\dddot{H}$.

\

\section{Calculation of the $\beta$-Bogoliubov coefficient} \label{appB}

For the potential (\ref{PV}), we use the second iteration of the WKB solution and we write 
 $\chi_{x,WKB}^{(2)}= \frac{1}{\sqrt{2W_k^{(2)}}}e^{-i\int^{\tau}W_k^{(2)}(\eta)d\eta}$. Then,  before the phase transition the positive frequency mode evolves approximately as 
$\chi_{x,WKB}^{(2)}$ but after the abrupt phase transition the positive and negative frequencies mix and the mode evolves approximately as 
$\alpha_k\chi_{x,WKB}^{(2)}+\beta_k (\chi_{x,WKB}^{(2)})^*$. By matching both expressions at the beginning of the kination  phase, which we have assumed to be at $\varphi=0$ as we have already explained (see also Figure \ref{retrat}), we obtain 
\begin{eqnarray}
\beta_k=\frac{\mbox{W}[\chi_{x,WKB}^{(2)}(\tau_{kin}^+); \chi_{x,WKB}^{(2)}(\tau_{kin}^-) ]}
{\mbox{W}[\chi_{x,WKB}^{(2)}(\tau_{kin}^+); (\chi_{x,WKB}^{(2)})^*(\tau_{kin}^+) ]}
\cong 
-i{\mbox{W}[\chi_{x,WKB}^{(2)}(\tau_{kin}^+); \chi_{x,WKB}^{(2)}(\tau_{kin}^-) ]}
,
\end{eqnarray}
where $\tau_{kin}$ denotes the beginning of kination,
$\mbox{W}[f;g]=fg'-gf'$ is the wronskian of the functions $f$ and $g$,  and
$f(\tau_{kin}^{\pm})=\lim\limits_{\tau\rightarrow \tau_{kin}^{\pm}}f(\tau)$ denotes the values of $f$ immediately before and after the beginning of kination.

\

Now a simple calculation shows that
{\footnotesize
\begin{eqnarray}\label{10}
|\beta_k|^2=\frac{1}{4W_k^{(2)}(\tau_{kin}^+)    W_k^{(2)}(\tau_{kin}^-)}\left[(W_k^{(2)}(\tau_{kin}^+)-W_k^{(2)}(\tau_{kin}^-))^2
+\frac{1}{4}
\left( \frac{W^{(2)\prime}_k(\tau_{kin}^+)}{W_k^{(2)}(\tau_{kin}^+)}- \frac{W^{(2)\prime}_k(\tau_{kin}^-)}{W_k^{(2)}(\tau_{kin}^-)}  \right)^2\right],
\end{eqnarray}}
which for our model can be approximated by 
\begin{eqnarray}
|\beta_k|^2=\frac{1}{4}\frac{(W_k^{(2)}(\tau_{kin}^+)-W_k^{(2)}(\tau_{kin}^-))^2}
{W_k^{(2)}(\tau_{kin}^+)    W_k^{(2)}(\tau_{kin}^-)    }
\cong
\frac{(W_k^{(2)}(\tau_{kin}^+)-W_k^{(2)}(\tau_{kin}^-))^2}
{4\omega_{k,kin}^2  }
,
\end{eqnarray}
where we have introduced the notation $\omega_{k,kin}\equiv \omega_k(\tau_{kin})$.

Then, using the leading terms of $W_k^{(2)}$, obtained in the Appendix A, we reach
\begin{eqnarray}
|\beta_k|^2\cong \left\{\begin{array}{cc}
 \frac{m_{\chi}^4 a_{kin}^{12}}{1024
 \omega^{12}_{k,kin}}(\dddot{H}(\tau_{kin}^+)-\dddot{H}(\tau_{kin}^-))^2& \quad  \mbox{for the conformally coupled case} \\
 & \\
 \frac{9a_{kin}^{8}(\xi-1/6)^2}{64\omega^{8}_{k,kin}}
 (\dddot{H}(\tau_{kin}^+)-\dddot{H}(\tau_{kin}^-))^2
 & \quad  \mbox{for the nonconformally coupled case,} \end{array} \right.
 \end{eqnarray}
and, in order to obtain the value of $\dddot{H}(\tau_{kin}^+)-\dddot{H}(\tau_{kin}^-)$,
first of all 
we take the time derivative of the conservation equation, namely 
$
\dddot{\varphi}+3\dot{H}\dot{\varphi}+ 
3{H}\ddot{\varphi}+V_{\varphi\varphi}
\dot{\varphi}=0,
$
which leads to
\begin{eqnarray}
\dddot{\varphi}(\tau_{kin}^+)-
\dddot{\varphi}(\tau_{kin}^-)=-\dot{\varphi}_{kin}(
V_{\varphi\varphi}(0^+)-V_{\varphi\varphi}(0^-)),
\end{eqnarray}
where we have used that up to the second derivative the scalar field $\varphi$ is continuous at the beginning of kination.

\

On the other hand, from Raychaudhuri equation
$\dddot{H}=-\frac{1}{M_{pl}^2}(\ddot{\varphi}^2+\dot{\varphi}\dddot{\varphi})$, we get
\begin{eqnarray}
\dddot{H}(\tau_{kin}^+)-\dddot{H}(\tau_{kin}^-)=\frac{\dot{\varphi}^2_{kin}}{M_{pl}^2}(
V_{\varphi\varphi}(0^+)-V_{\varphi\varphi}(0^-))=-\frac{4\lambda}{3}
\dot{\varphi}_{kin}^2,
\end{eqnarray}
which finally leads to
 \cite{haro19}
\begin{eqnarray}\label{beta}
|\beta_k|^2\cong \left\{\begin{array}{cc}
 \frac{m_{\chi}^4 \lambda^2a_{kin}^{12}\dot{\varphi}_{kin}^4}{576\omega^{12}_k(\tau_{kin})}& \quad  \mbox{for the conformally coupled case} \\
 & \\
 \frac{\lambda^2a_{kin}^{8}\dot{\varphi}_{kin}^4}{4\omega^{8}_k(\tau_{kin})}(\xi-1/6)^2& \quad  \mbox{for the nonconformally coupled case.} \end{array} \right.
 \end{eqnarray}

 \
 
 Finally, dealing with the potential (\ref{PV1}), we only have to use the first order WKB solution, 
 and then the equation (\ref{10}), changing  $W_k^{(2)}$ by  $W_k^{(1)}$, can be approximated by
\begin{eqnarray}
|\beta_k|^2\cong \frac{1}{16\omega_{k,kin}^4}
\left( {W_k^{(1)\prime}}(\tau_{kin}^+)- {W_k^{(1)\prime}}(\tau_{kin}^-)  \right)^2.
\end{eqnarray}
Now, taking into account that for the conformally coupled case the leading term of $W_k^{(1)}$ is 
$-\frac{1}{4\omega_k}\frac{\omega_k''}{\omega_k}$ and for the nonconformally coupled one is 
$\frac{1}{2\omega_k}(\xi-1/6)a^2R$ (see formula (\ref{first})), after some algebra we obtain
\begin{eqnarray}
|\beta_k|^2\cong\left\{\begin{array}{ll}\frac{m_{\chi}^4a_{kin}^{10}}{256\omega_{k,kin}^{10}}(\ddot{H}(\tau_{kin}^{-})-\ddot{H}(\tau_{kin}^{+}))^2 &  \mbox{for $\xi=1/6$,} \\
& \\
\frac{9a_{kin}^6}{16\omega_{k,kin}^6}(\ddot{H}(\tau_{kin}^{-})-\ddot{H}(\tau_{kin}^{+}))^2 & \mbox{for $|\xi-\frac{1}{6}|\cong 1$,} \end{array}\right.
\end{eqnarray}
where,  { from conservation and Raychaudhuri equations, we have that} $(\ddot{H}(\tau_{kin}^{-})-\ddot{H}(\tau_{kin}^{+}))^2=
\dot{\varphi}_{kin}^2\left(\frac{V_{\varphi}(0^{-})}{M_{pl}^2} \right)^2
{=\frac{2\lambda^2}{3}
\dot{\varphi}_{kin}^2 M_{pl}^2 }$  and, thus,

\begin{eqnarray}\label{beta1}
|\beta_k|^2\cong\left\{\begin{array}{ll}
\frac{m_{\chi}^4a_{kin}^{10}\lambda^2}
{384\omega_{k,kin}^{10}}\dot{\varphi}_{kin}^2 M_{pl}^2 
 &  \mbox{for $\xi=1/6$,} \\
& \\
\frac{3a_{kin}^6\lambda^2}{8\omega_{k,kin}^6}
\dot{\varphi}_{kin}^2 M_{pl}^2 
 & \mbox{for $|\xi-\frac{1}{6}|\cong 1$.} \end{array}\right.
\end{eqnarray}

  \end{appendices}
 
}

  }

\end{document}